\documentclass{aa}
\usepackage{graphicx}
\usepackage{txfonts}
\usepackage{natbib}
\usepackage{rotating}

\def\Hubble{$\rm \,km\,s^{-1}\,Mpc^{-1}$}
\def\H2{H$_2$}
\def\msol{M$_{\odot}$}

\def\micro{$\rm \,\mu m$}

\def\Tim{TIMMI\,2}
\def\Xray{X-ray}
\def\Xrays{X-rays}

\begin{document} 

\title{Mid-Infrared Imaging of Active Galaxies}
\subtitle{Active Nuclei and Embedded Star Clusters}

\author{E. Galliano \inst{1}
	\and D. Alloin \inst{1,2}
	\and E. Pantin \inst{2}
	\and P.O. Lagage \inst{2}
	\and O. Marco \inst{1}}

\institute{European Southern Observatory, Casilla 19001, Santiago 19, Chile
	\and UMR7158 CEA-CNRS-U.Paris7, DSM/DAPNIA/Service d'Astrophysique, CEA/Saclay, 91191 Gif-sur-Yvette, France}

\offprints{E. Galliano,
\email{egallian@eso.org}}

\date{Received / Accepted }

\titlerunning{MIR imaging of galactic nuclear regions}

\abstract{High resolution, mid-infrared (MIR) images of a set of nine nearby active galaxies are presented. The data were obtained with the \Tim~instrument mounted at the ESO 3.6\,m telescope using a set of N-band narrow filters. The resulting images have an angular resolution of 0.6\arcsec-1\arcsec. The MIR emission has been resolved in four galaxies: NGC253, NGC1365, NGC1808 and NGC7469. The images unveil a circumnuclear population of unknown MIR sources in NGC1365 and NGC1808, coincident with radio sources. These MIR/radio sources are interpreted in terms of embedded young star clusters. A high-resolution MIR map of NGC253 is also presented, and enables the identification of a previously unknown MIR counterpart to the radio nucleus. Extended MIR emission is detected in NGC7469, and concurs with previous observations in the NIR and radio. For this source, an interesting morphological difference between the 10.4\,\micro~and the 11.9\,\micro~emission is observed, suggesting the presence of a dust-rich micro-bar. Our MIR images of Circinus do not show resolved emission from the nucleus down to an angular scale of 0.5\arcsec. In the case of NGC2992, an upper limit to the extended MIR emission can be set. Finally, we provide new MIR flux measurements for the unresolved AGN in NGC5995, IZw1 and IIZw136.
	}
	\authorrunning{Galliano et al.}
	\titlerunning{MIR imaging of active galaxies}
\maketitle

\section{Introduction}
\label{introduction}

As mid-infrared (MIR) radiation is much less affected by extinction than visible light, dust embedded objects in the nuclear regions of galaxies are more efficiently detected in this wavelength range. High spatial resolution in the MIR requires the use of large-aperture telescopes and thus is restricted to ground-based instruments that provide a substantial gain in spatial resolution over currently available MIR instruments in space. The gain in spatial resolution is, in the context of active galaxies, essential since in many cases the relative contributions to their total power of the starburst activity and of the active galactic nucleus (AGN) are unknown. Let us briefly discuss how high resolution MIR observations can help probe/disentangle these two phenomena.

Young bursts of star formation are likely to be substantially obscured by dust. Therefore, high resolution MIR imaging is an appropriate means of identifying them. Indeed, recent infrared (IR) observations have revealed the existence of very bright embedded star clusters, containing few thousands O stars, which are invisible on images obtained in the optical. The first object of this type to be discovered was the deeply buried star cluster in the Antennae galaxy by \citet{Mirabel98}. It produces 20\% of the total MIR emission of the galaxy, and contains $1.6 \times 10^7$\,\msol~of stars hidden behind an $A_V=10$ cloud of dust \citep{Gilbert00}. Similar objects have been found in NGC5253 \citep{Gorjian01}, SBS 0335-052 \citep{Plante02} and IIZw40 \citep{Beck02}.

High resolution MIR imaging is also essential for the identification and classification of AGN. For sources affected by intrinsically large extinction, an optically-based classification can be quite misleading. For example, \citet{Krabbe01} noted that the classification of ``active galaxies'' based solely upon their optical properties is a poor indicator since \Xray~or radio observations could reveal the presence of an AGN in ``classical'' starburst galaxies. Conversely, ISO observations \citep{Laurent99} of optically classified AGN show that their MIR emission is often dominated by star forming regions rather than by the AGN itself. Thus by spatially resolving emission from star forming regions and AGN, and photometrically measuring each through high resolution imaging, one can determine their relative contributions to the total integrated power. For AGN, precise and spatially resolved MIR photometry is mandatory in order to construct spectral energy distributions (SED) that are free from starburst contributions. The resultant SED may then be compared to models developed in the framework of the unified scheme. Due to the intrinsic small size of the dusty torus in AGN (1 to a few $\times$100\,pc), only a few can be directly probed through a spatially resolved mapping of their MIR emission. 
One of these is the extensively studied type 2 AGN in NGC1068. MIR imaging of this source at 0.6\arcsec~resolution revealed extended and collimated 10 and 20\,\micro~emission up to 5\arcsec~from the core, along the direction of the radio jet \citep{Alloin00}. At higher angular resolution \citep[0.1\arcsec~after deconvolution:~][]{Bock00,Tomono01}, the complex structure of the MIR emission is found to be coincident with the narrow line region (NLR) clouds \citep{Galliano03b}. Therefore, the access to high resolution imaging brings new clues regarding the morphology of the material around the central engines of AGN.  

This paper presents high angular resolution MIR images of a sample of active galaxies, obtained with the \Tim~instrument at the ESO 3.6\,m telescope, La Silla observatory. Observations and data reduction are presented and discussed in Sect.~\ref{observations and data reduction}~and the results are given in Sect.~\ref{results}. First, the general results such as flux measurements and maps for the resolved objects are given in Sect.~\ref{general results}. Then, we provide more detailed analyses for NGC253 (Sect.~\ref{NGC253}), Circinus (Sect.~\ref{Circinus}), NGC1808 (Sect.~\ref{sect:NGC1808}), NGC1365 (Sect.~\ref{sect:NGC1365}) and NGC7469 (Sect.~\ref{NGC7469}). Sect.~\ref{discussion1} gathers high spatial resolution IR flux measurements and provides SED for the targets. Finally, we discuss in Sect.~\ref{discussion2} the newly discovered populations of MIR sources in NGC1808 and NGC1365: a simple model supports the interpretation in terms of embedded star clusters. A detailed discussion of these sources will be presented along with new observations in a forthcoming paper. Summary and conclusions are given in Sect~\ref{conclusion}. A Hubble constant $H_0=75$\Hubble~is used throughout the article to compute the distance to the objects, except when measured with a more precise method.

\section{Observations, data reduction and templates}
\label{observations and data reduction}
\subsection{Observations}
\label{observations}
\begin{table*}[htbp]
\caption[]{General properties of the sample galaxies: coordinates, distance, AGN type, IR flux}
\begin{center}
\begin{tabular}{|lllllll|} \hline 
Galaxy & RA    & Decl.    & Dist. & Scale           & Activity    & IRAS 12\micro~flux$^c$   \\
       & (J2000) & (J2000) &  Mpc        & pc/arcsec &        &   Jy \\ \hline 
NGC253$^a$ & 00$^{\rm h}$47$^{\rm m}$33.1$^{\rm s}$ & -25\degr17\arcmin18\arcsec & $2.5$ & 12 & Starburst + AGN?   & 24.0\\
Circinus & 14$^{\rm h}$13$^{\rm m}$09.3$^{\rm s}$ & -65\degr20\arcmin21\arcsec & $4.0$ & 19 & Type\,2 AGN   & 18.8 \\
NGC1808$^b$ &  05$^{\rm h}$07$^{\rm m}$42.3$^{\rm s}$ & -37\degr30\arcmin46\arcsec & $10.9$ & 53 & Starburst + AGN? & 4.4  \\ 
NGC1365 & 03$^{\rm h}$33$^{\rm m}$36.4$^{\rm s}$ & -36\degr08\arcmin25\arcsec & $18.6$ & 90 & Type\,2 AGN   & 3.4 \\
NGC2992  & 21$^{\rm h}$32$^{\rm m}$27.8$^{\rm s}$  & +10\degr08\arcmin19\arcsec & $30.1$ & 150 & Type\,2 AGN  &  -- \\
NGC7469 & 23$^{\rm h}$03$^{\rm m}$15.6$^{\rm s}$ &   +08\degr52\arcmin26\arcsec & $66$ & 320 & Type\,1 AGN & 1.35\\
NGC5995 & 15$^{\rm h}$48$^{\rm m}$24.9$^{\rm s}$ & -13\degr45\arcmin28\arcsec & $100$ & 490 & Type\,2 AGN & 0.39 \\
IZw1 & 00$^{\rm h}$53$^{\rm m}$34.9$^{\rm s}$ & +12\degr41\arcmin36\arcsec & $245$ & 1186 & Type\,1 AGN  & 0.51 \\
IIZw136 & 21$^{\rm h}$32$^{\rm m}$27.8$^{\rm s}$ &  +10\degr08\arcmin19\arcsec & $252$ & 1222 & Type\,1 AGN  & $<$0.2 \\ \hline 

\end{tabular}
\end{center}
$^a$ A discussion about the nature of the radio nucleus of NGC253 can be found in \citet{Forbes00}.

$^b$ The AGN nature of the nucleus of NGC1808 is not clearly established, see the discussion in \citet{Krabbe01} and in Sect.~\ref{discussion1}

$^c$ Flux of the entire galaxy, from \citet{Moshir90}.

\label{obs_summary} 
\end{table*}

The original aims of this study were two-fold: (1) perform preliminary observations of active galaxies in order to prepare future observations with VISIR at the VLT, (2) look for extended MIR emission directly related to the AGN, as was observed along the radio jet direction in NGC1068 \citep{Alloin00,Bock00,Tomono01}. To address these aims, we chose a sample of nearby (few Mpc) to moderately nearby (few hundreds of Mpc) AGN. The observations quickly showed interesting by-products such as the star forming environment around some AGN: the MIR counterparts of known radio sources in NGC1365 and NGC1808 became unveiled. After these first findings, we decided to include in our sample NGC253, the archetypal starburst galaxy. Table~\ref{obs_summary} displays basic information about the targets.

The observations were performed with \Tim~(Thermal Infrared Multi Mode Instrument\footnote{http://www.ls.eso.org/lasilla/Telescopes/360cat/timmi/}) installed at the 3.6m telescope on La Silla during four runs, on 2001 March 10-11, July 5, October 8 and November 26.  \Tim~is equipped with a 240 $\times$ 320 pixel AsSi BIB detector and we observed with an 0.2\arcsec~pixel size, resulting in a field of view of 48\arcsec $\times$ 64\arcsec.

Four narrow filters spanning the N-band were available: the 8.9\micro~filter (7.90-9.46\micro), the 10.4\micro~filter (9.46-11.21\micro), the 11.9\micro~filter (10.61-12.50\micro) and the 12.9\micro~ filter (11.54-12.98\micro). The labels of the filters do not correspond to their effective wavelengths, but rather refer to the spectral features they probe. One filter is also available in the Q-band (20\micro). The seeing was good, $\sim$0.6-1.0\arcsec~in the visible and allowed to obtain diffraction-limited images at all observed MIR wavelengths. The main difficulties in performing MIR observations from the ground are the high level of sky emission and the variability of the sky transparency, particularly in the 20\micro~window. 

\subsection{Data reduction}
\label{data reduction}

The standard chopping/nodding technique was used to determine the sky and telescope background for subtraction. We observed in ``small imaging mode'', which means that the nodding direction was perpendicular to the chopping direction. On-line reduction at the telescope allowed the chopping throws and nodding offsets to be adjusted depending on the extension of the sources. Chopping throws between 15\arcsec~and 20\arcsec~and nodding offsets from 15\arcsec~to 30\arcsec~were used. MIR standard stars were observed before and after each source observation in order to calibrate the photometry of the final images, evaluate variations of the sky transparency, and estimate the Point Spread Functions (PSF) for further image deconvolution. \texttt{IDL} routines were written and used for data reduction. Glitches were carefully removed and the data were corrected for the \Tim~bias: each of the sixteen readout channels of the chip has a response which varies with time. This appears on the images as sixteen vertical bands with slightly different intensity levels. This detector feature makes the flat field correction useless, as it assumes a stable detector response, at least over the time scale of an observing night. These reduction steps were performed on each plane of the data-cubes, each of these planes corresponding to the subtraction of two consecutively chopped images. Eventually, the planes of each reduced data-cube were co-added resulting in the scientifically exploitable image.

Where resolved structures were present, we have deconvolved the images in order to better separate the different sources. Deconvolution was performed using the \texttt{mr\_deconv} routine of the \texttt{MR/1} software package developed by \citet{Murtagh99}. Standard stars were used to determine the PSF. The \texttt{MR/1} software package performs an efficient regularisation in the wavelet space. This procedure is excellent at pulling low signal out of noise, clarifying the distribution of sources. The results obtained after deconvolution were suitable for comparison with high resolution cm radio images and we could indeed clearly identify MIR counterparts for a number of radio sources. For low brightness sources within the resolved structures, we measured the fluxes on the deconvolved images. The precision of this procedure was tested with the following simulation: an artificial MIR image of NGC1808 was created, in which weak point sources were given the fluxes measured on the deconvolved image. Poisson noise corresponding to the high background level was added. Then, this fake image was deconvolved: after deconvolution, the input fluxes were recovered with an accuracy better than 10\%. The brightest among the weak sources were all recovered, while a few false very weak sources appeared on the deconvolved image, but could be easily identified and excluded since they were more compact than real sources. 

\subsection{Photometric error estimation}
\label{errors}
The uncertainty on the flux measurements is dominated by variations of the sky transparency. In the 10\micro~window, variations of the photometric calibration (ADU to flux) during the night were found to be of about 10\%. One extreme case of transparency variation occured on 2001 November 26: the photometric calibration varied by 50$\%$ in about 20\,min probably because of spatial variations of the sky transparency. This happened during the observing sequence for NGC1365 through the 12.9\micro~filter. The corresponding data-cube was checked: we split it into equal slices, for each of which the source fluxes were measured and found to be equal. This ensured that the sky transparency had not changed dramatically while performing the science observation, but rather differed between the two standard star observations. Unfortunately, there is no possibility to check the spatial variation of the sky transparency. In conclusion, we will keep in mind that the 12.9~\micro~flux measured for NGC1365 is uncertain. One must also consider that the values for the standard star fluxes are known to a precision that is not better than 10\%. Finally, we adopt a formal error of $\pm20\%$ on the quoted flux measurements. 

\subsection{Templates for MIR sources: AGN, HII~region and PDR}
\label{MIR emission}
\begin{figure}[htbp]
\begin{center}
\resizebox{8cm}{!}{\includegraphics*[scale=1.]{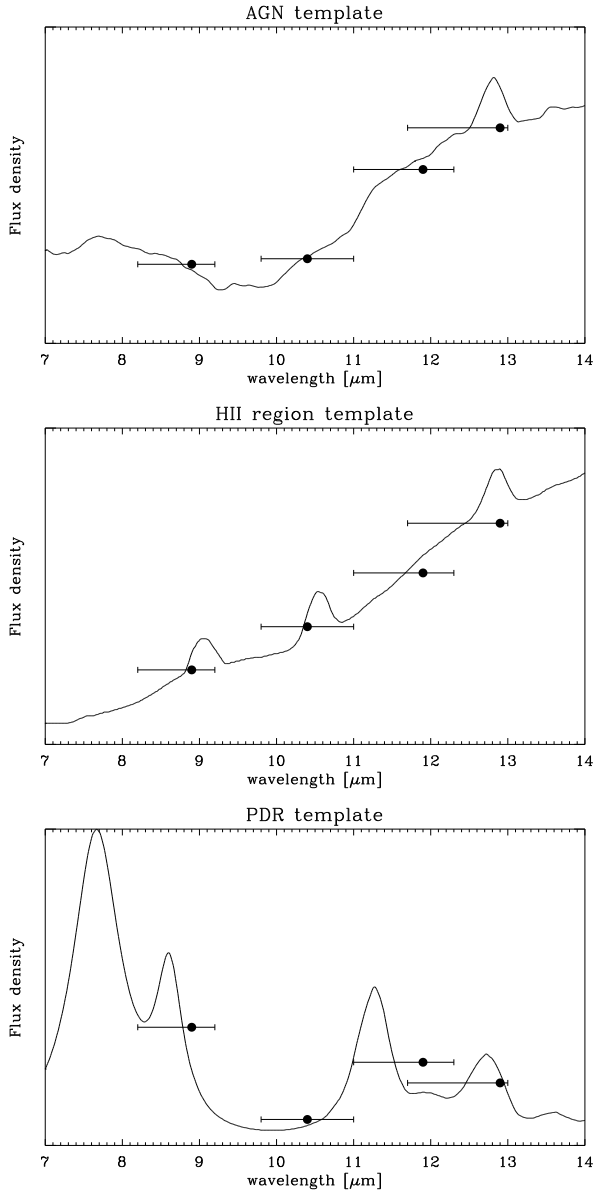}}
\caption{Template spectra for the 7-14\micro~emission of the three main types of MIR sources: AGN, HII~region and PDR. The spectra are from \citet{Laurent00}. The corresponding flux densities through the 8.9\micro, 10.4\micro, 11.9\micro~and 12.9\micro~filters are overplotted: the horizontal bar marks the width of the filter, while the abscissa of the dots correspond to the wavelengths of the \Tim~filter names.} 
\label{templates}
\end{center}
\end{figure}

The MIR emission observed in galaxies has different origins: 

\begin{itemize}

\item The Rayleigh-Jeans emission of the evolved stellar population is the dominant contribution in early-type galaxies which are poor in gas and dust.

\item The dust, irradiated and heated by an underlying continuum, emits a MIR continuum. Large grains are in thermal equilibrium with the radiation field and have a blackbody-like emission. There also exists a population of very small grains (size $\leq$ 50\AA) which never reach the thermal equilibrium. These very small grains (hereafter VSG) can get to high temperatures (T $>$ 100K) after absorbing one single UV photon. The grains cool down by emission longwards of 10\micro~before they absorb new photons. The VSG emission is typical of regions which are actively forming stars, such as HII~regions.

\item The ionised gas of the interstellar medium emits several strong forbidden lines in the MIR. Lines from Ne, Ar, Si as well as S and Fe are commonly observed in the spectra of AGN (excluding BLLac and OVV) and HII~regions. In particular, [ArIII] 8.9\micro, [SiIV] 10.4\micro~and [NeII] 12.8\micro~lie within the wavelength range of the set of filters we used for this study. 
  
\item The unidentified infrared bands (UIB) observed at 6.2, 7.7, 8.6, 11.3 and 12.7\micro~are characteristic of photo-dissociation regions (hereafter PDR), where molecules form and their physical conditions depend on the UV photon density. Since the original work of \citet{Leger84}, these bands are interpreted as the bending and stretching vibration modes of the C=C and C-H links in polycyclic aromatic hydrocarbons (PAH).   

\end{itemize}

In Fig.~\ref{templates}, template spectra over the 7-14\micro~range \citep{Laurent00} are displayed for the three main types of MIR sources: AGN, HII~region and PDR. These templates are useful for comparison with observations: we therefore overplotted on the template spectra the flux densities that would be measured through the \Tim~filters.

The AGN spectrum is dominated by continuum emission from the dusty torus, and hard silicate absorption around 9.7\micro~can be observed. Usually, PAH emission is absent or very weak since the carriers are destroyed by the strong radiation field.

The HII~region spectrum displays a steeply rising continuum, due to VSG emission. The ionised gas produces the strong [ArIII] 8.9\micro, [SiIV] 10.4\micro~and [NeII] 12.8\micro, but the PAH carriers are destroyed by the radiation field as in the case of AGN.

The PDR spectrum is dominated by the emission of the PAH. The PDR forms at a distance from the hot stars which is large enough for the UV continuum to be weakened and softened; this allows molecules to be formed in the cool gas. 

We shall use the flux ratios computed with these templates to distinguish between the various object types and to help interpret the MIR emission from our sample sources.

\section{Results}
\label{results}

\begin{table*}[h]
\caption[]{Photometry of resolved MIR sources for each galaxy in our sample in each N-band filter. The symbol '--' indicates that the source was not observed through the corresponding filter. Source labelling is according to the scheme described in Sect.~\ref{results}.}
 \begin{center}
\begin{tabular}{|lc|cccc|cc|} \hline
Target		&Individual	&Flux [mJy]$^a$	&Flux [mJy]$^a$	&Flux[mJy]$^a$	&Flux[mJy]$^a$	& \multicolumn{2}{c|}{Relative position$^b$} \\ 
 &		source name	&8.9\micro	&10.4\micro	&11.9\micro	&12.9\micro	& $\alpha$ [\arcsec]	& $\delta$ [\arcsec]    		\\ \hline 
NGC253		&	M1	&       --	&	--	&$5100\pm1000$	&	--	&0 & 0 		\\

		&	M2	&	--	&	--	&$400\pm100$	&	--	&0.87 & 1.04 		\\ 

		&	M3	&	--	&	--	&$350\pm100$	&	--	&1.17 & 1.5               \\

		&	M4	&	--	&	--	&$500\pm100$	&	--	&1.99 & 1.29            \\

		&	M5	&	--	&	--	&$250\pm50$	&	--	&2.73 & 2.81  	\\

		&	M6	&	--	&	--	&$250\pm50$	&	--	&4.15 & 3.68 \\

		&$\sigma$ bg	&	--	&	--	&60		&	--	&	& 	\\ \hline
Circinus$^c$	&	M1	&$7950\pm1600$	&$6650\pm1350$	&$16650\pm3300$	&$23400\pm4700$	& & 		\\ 

		&$\sigma$ bg	&50		&	25	&20		&25		&	& 	\\ \hline
NGC1808		&	M1	&	--	&$255\pm50 $    &$620\pm120$  	&$970\pm190$ 	& 0&0           \\
						                                                
		&	M2	&	--	&$\leq 10 $     & $40 \pm  5 $  &$45 \pm5 $ 	&-1.99  & -3.76  \\
						                                                 		 
		&	M3	&	--	&$10\pm2$    	&$50\pm10 $  	& $80 \pm10 $ 	&-4.55  &  1.39  \\
						                                                 		 
		&	M4	&	--	&$ \leq 10 $    & $70 \pm10 $   &$170 \pm30 $ 	& 2.77  &  0.54  \\
						                                                 		 
		&	M5	&	--	&$23 \pm5$      & $\leq 50  $   & $90 \pm 10 $  & 2.87  & -1.59  \\
						                                                 		 
		&	M6	&	--	&$\leq 10 $     &$75\pm10$  	&$145 \pm20 $ 	& 5.82  & -5.25  \\
						                                                 		 
		&	M7	&	--	&$40 \pm    5 $ & $125 \pm20 $ & $90\pm20  $ 	& 5.44  & -7.34  \\
						                                                 		 
		&	M8	&	--	&$15 \pm    5 $ & $105 \pm20 $ & $195 \pm30 $ 	& 2.71  & -5.51  \\
						                                                 		 
		&	M9	&	--	&$30 \pm   5 $  & $\leq 40$    & $60\pm10 $  	& 0.70  & -3.38  \\
 
		&$\sigma$ bg     &	--	&	10	&20		&25		&	& 	\\ \hline
NGC1365		&	M1	&$410\pm80$     &$440\pm 80$  	&$510 \pm 100$ &$1100 \pm 220^d$	&0&0\\
				                                                                
		&	M2	&$\leq40$       &$ 20\pm  4$  	&$\leq 20$    	&$\leq 40$  	& -4.66  & -5.12\\  
				                                                                  		 
		&	M3	&$\leq 40$      &$ 30\pm  6$  	&$ 30 \pm  6$  	&$\leq 25$ 	& -5.40  & -2.56 \\ 
				                                                                  		 
		&	M4	&$\leq 40$      &$\leq 5$   	&$ 40 \pm  8$ &$160 \pm  30^a$	& 0.41  &  7.12  \\
				                                                                  		 
		&	M5	&$\leq 40$      &$ 60\pm 10$  	&$120 \pm 20$  &$320 \pm  60^a$	& 2.79  &  9.99 \\
				                                                                  		 
		&	M6	&$105\pm20$     &$ 50\pm 10$  	&$140 \pm 20$  &$505 \pm 100^a$	& 4.75  &  6.95  \\

		&$\sigma$ bg	&35		&	10	&	10	&	20	&	& 	\\ \hline
NGC2992		&	M1	&	--	&$230\pm45$	&	--	&$565\pm110$	&&		\\ 

		&$\sigma$ bg	&	--	&18		&	--	&33		&&		\\ \hline
NGC7469		&\diameter =$2\arcsec$	&	--	&$310\pm60$	&$565\pm110$&	--	&&		\\ 

		&\diameter =$5.6\arcsec$	&	--&$450\pm90$&$980\pm200$&	--	&&		\\ 

		&$\sigma$ bg	&	--	&	10	&	10	&	--	&&		\\ \hline
NGC5995		&	M1	&$310\pm60$	&	--	&$300\pm60$	&	--	&&		\\ 

		&$\sigma$ bg	&100.		&	--	&25.		&	--	&&		\\ \hline
IZw1		&	M1	&	--	&$375\pm75$	&$425\pm85$	&$325\pm65$	&&		\\ 

		&$\sigma$ bg	&	--	&10		&18		&16		&&		\\ \hline
IIZw136		&	M1	&	--	&$130\pm25$	&$130\pm25$	&	--	&&		\\

		&$\sigma$ bg	&	--	&	13	&	13	&	--	&&		\\ \hline
\end{tabular}
\end{center}

None of the individual sources from this table are resolved on the \Tim~images, except tentatively NGC1365/M1 (see Sect.~\ref{sect:NGC1365}). The linear upper limits for the source sizes are, given the galaxy distances of Table~\ref{obs_summary}:

NGC253: 12\,pc; Circinus: 7\,pc; NGC1808: 40\,pc; NGC1365: 70\,pc; NGC2992:  120\,pc; NGC7469: 250\,pc for the central peak; NGC5995: 400\,pc; IZw1 and IIZw136: 1\,kpc.

$(^a)$ Unit for the fluxes is mJy and the unit for $\sigma$ bg is mJy\,arcsec$^{-2}$. For most sources, the quoted error is the 20\%~calibration error (see Sect.~\ref{errors}) which dominates over the measurement uncertainty on the images. In cases where the measurement uncertainty  was not negligible compared to the 20\%~error, it was included.

$(^b)$ The positions are offsets on our MIR images, relative to source M1, the coordinates of which are set to (0\arcsec,0\arcsec). No significant astrometry is possible with this \Tim~data set because of the small field of view. For NGC253, a discussion on the coordinates of M1 (corresponding to the 0\arcsec,0\arcsec~position in this table) is provided in Sect.~\ref{NGC253}. For NGC1808, M1 can be identified to its counterpart on the cm image (see Sect~\ref{sect:NGC1808}), The B1950 coordinates of the M1 cm counterpart derived by \citet{Saikia90} are RA: 05$^{\rm h}$05$^{\rm m}$58.56$^{\rm s}$, Dec:-37\degr34\arcmin36.3\arcsec, with no information concerning the uncertainty. For NGC1365, the B1950 coordinates of M1 radio counterpart are RA: 03$^{\rm h}$31$^{\rm m}$41.80$^{\rm s}$, Dec:-36\degr18\arcmin26.8\arcsec \citep[][ no information on the uncertainty]{Sandqvist95}. This cm image is shown in Sect.~\ref{sect:NGC1365}.  
 
$(^c)$ Circinus was also detected at 20\micro, with a flux of $30\pm6\,\rm Jy$ and $\sigma$ bg of 1.2\,Jy\,arcsec$^{-2}$.  

$(^d)$ The photometric calibration used for NGC1365 at 12.9\micro~may be greater than the actual one by up to a factor two because of local sky transparency variations (see Sect.~\ref{errors}). This would result in flux values that are up to twice smaller.

\label{main table} 
\end{table*}
\begin{figure*}[htbp]
\begin{center}
\resizebox{18cm}{!}{\includegraphics*[scale=1.]{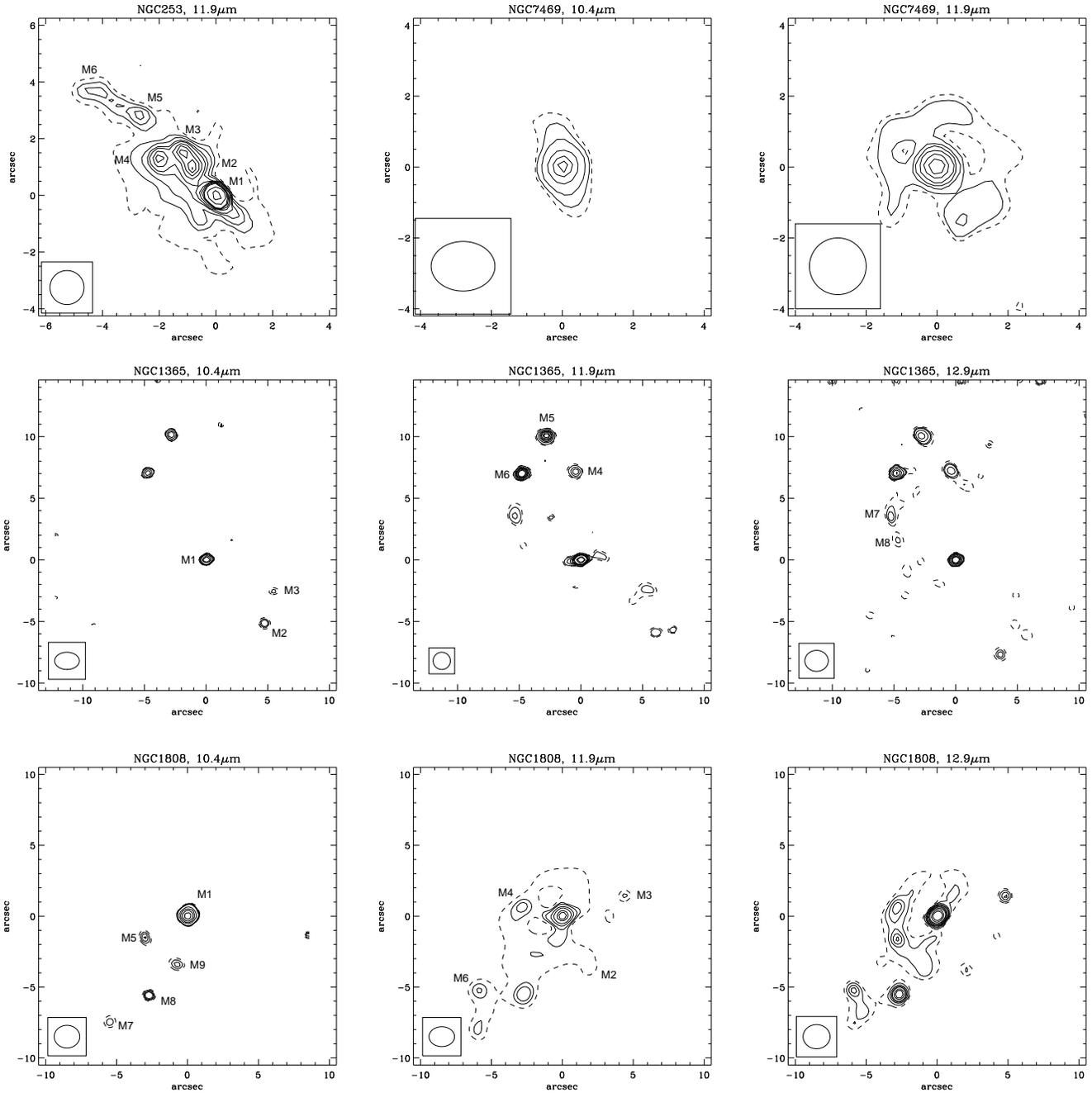}}
\caption{Deconvolved maps for the objects with resolved emission. North is up and East left. Dotted contours correspond to 1$\sigma$ of the background. Source labelling is according to the scheme described in Sect.~\ref{results}. Galaxy and filter identifications are given as a title for each panel. The lower left square on each panel displays the corresponding PSF shape (FWHM in the E-W and N-S directions).}
\label{All contours}
\end{center}
\end{figure*} 
An overview of the observational results is  presented in this section. In the first part, we give results for the whole sample, and provide a short general discussion. Then we discuss in more detail the most interesting cases. Table~\ref{main table} summarises the full set of measurements. For each target, we provide the fluxes and relative positions for all the identified sources. The targets are listed by increasing distance. MIR extended emission (resolved and/or diffuse) is detected in 4 objects: NGC253, NGC1808, NGC1365 and NGC7469. Contour maps of the emission for these sources are presented in Fig.~\ref{All contours}. For each object, the brightest source (which is the AGN except in the case of NGC253) is labelled M1, while the circumnuclear sources are labelled M2, M3, etc...

\subsection{Overview}
\label{general results}

MIR emission was detected for all the targets in the sample: 

\begin{itemize}
\item In NGC253, six distinct sources have been identified in the central region. This provides a gain in spatial resolution with respect to previous observations, which helps clarify the distribution of the MIR emission from this galaxy.   
\item In Circinus, emission has been detected in all four narrow N-band filters. This is also the only target that we have detected at 20\micro. 
\item In NGC1808 and NGC1365, a previously unknown population of MIR sources in the inner 10\arcsec~region around their AGN is unveiled.  
\item In NGC7469, extended diffuse emission is detected around the AGN and shows interesting morphological peculiarities. While the 12.9\micro~map shows an emission quite symmetrically distributed around the nucleus, the 10.4\micro~map unveils the presence of a bar-like structure along the N-S direction.
\item The last four objects, NGC1992, NGC5995, IZw1 and IIZw136 are detected, each through two to three filters, as single unresolved sources.
\end{itemize}

The measured fluxes are generally in good agreement with previous measurements (see dedicated sections hereafter). Most previous measurements had been derived from data obtained with poorer spatial resolution and from photometry performed through much larger apertures than ours, hence encompassing a much larger fraction of the galaxies than the nuclear regions we are probing with \Tim. Therefore, this suggests that most of the MIR emission from these targets is indeed arising in their central $\sim$10\arcsec~region and can be accounted for by the AGN and the discrete sources revealed in the \Tim images. 

In none of the targets do we find extended MIR emission comparable to that observed in NGC1068 or 3C 66B along the radio jet \citep{Alloin00,Tansley00}, except a tentative detection in NGC1365 (see Sect.~\ref{sect:NGC1365}). According to the AGN unified scheme, we were expecting to detect collimated MIR emission in the type 2 objects of our sample. It is not so. Still, we believe that an analysis on a larger sample and from images with higher sensitivity is mandatory to understand the diversity of MIR morphologies of type 2 objects, and is left for future study. 

In the following, we present, in some detail, the results obtained for NGC253, Circinus, NGC1808, NGC1365, NGC7469. 

\subsection{NGC253: MIR detection of the radio nucleus?}
\label{NGC253}

\begin{figure}[htbp]
\begin{center}
\resizebox{7.5cm}{!}{\includegraphics*[scale=1.]{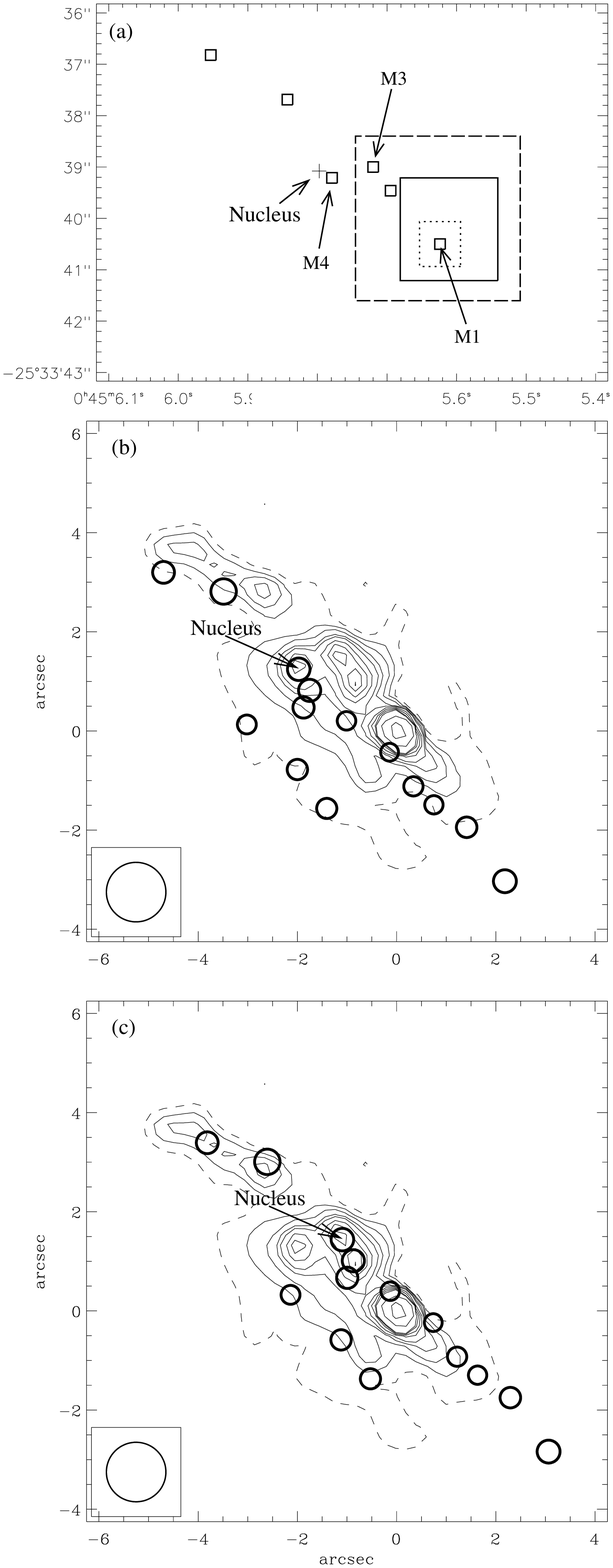}}
\caption{NGC253, two tentative registrations of the MIR map (contours) with respect to the radio map (brightest sources shown by circles). The radio map was obtained by \citet{Ulvestad97} at 2\,cm. (a) Absolute astrometry of the NIR/MIR images. The three boxes show the absolute (B1950) positions and positional errors for the bright NIR/MIR source M1: the dashed, solid and dotted line boxes show the respective registrations of \citet{Pina92}, \citet{Kalas94} and \citet{Keto93}. The small squares (0.2\arcsec~side) record the positions of the \Tim~sources with the brightest source M1 assigned at the most probable absolute position of peak1 as derived by \citet{Keto93}. The cross marks the absolute position of the radio nucleus. (b) Blind registration: M1 is positioned at the coordinates derived by \citet{Keto93}. (c) Best registration: tentative alignment of the MIR and radio peaks. As in Fig.~\ref{All contours}, the lower left square on each panel displays the PSF.}
\label{NGC253_timmi}
\end{center}
\end{figure} 

NGC253 hosts a vigorous starburst in its inner 100\,pc central region and is the archetypal starburst galaxy. \citet{Turner85} discovered a set of highly aligned compact radio sources along the $\sim$NE-SW direction. \citet{Ulvestad97}, performed high resolution and sensitive VLA observations and identified 64 sources. These sources probably belong to a nuclear ring seen at high inclination, and thus appear in a string-like distribution. The most powerful of these radio sources (TH-2) is considered to be the true nucleus of NGC253, and is thought to be an AGN \citep{Forbes00}. Imaging of NGC253 has been carried out in the NIR \citep{Forbes91,Forbes93,Sams94,Kalas94} and MIR \citep{Pina92,Keto93,Boeker98,Keto99}. These observations show that the NIR/MIR emission consists of a bright, barely elongated source plus a secondary, fainter source to the NE of the brightest one. The brightest NIR/MIR source is believed to be a young massive star cluster \citep{Keto99}. Both sources are inside an elongated envelope, with PA similar to that of the radio string. However, a general lack of correlation between the IR and the radio sources is observed. 

The \Tim~image of NGC253 (through the 11.9\micro~filter) is the highest resolution MIR image of this galaxy available to date. The MIR morphology revealed in this image is fully consistent with the previous MIR observations at lower angular resolution, but the distribution of the emission is resolved into six discrete sources. The flux measurement at 11.7\micro~made by \citet{Pina92} of M1 is best suited for comparison with our 11.9\micro~flux. Their flux of 4.8Jy is in good agreement with ours (5.1$\pm$0.1\,Jy). Comparison cannot be made with fluxes in the other filters because of the presence of strong emission lines and PAH bands \citep{Kalas94,Keto99}. 

Again, at the resolution achieved with \Tim, the source distributions in the radio and in the MIR are not obviously coincident, even though the global position angle of both distributions is the same. 

We propose two registrations of our MIR map with respect to the 2\,cm map by \citet{Ulvestad97}: 

(a) We use the absolute registration of M1 performed by \citet{Keto93}. It is consistent with other measurements from \citet{Pina92} and \citet{Kalas94}, and has the smallest uncertainty (about 0.4\arcsec): the measurement of the position of M1 gives $\alpha$(1950)=00$^{\rm h}$45$^{\rm m}$5.630$^{\rm s}\pm0.026^{\rm s} $ \& $\delta$(1950)=-25\degr33\arcmin40.57\arcsec$\pm0.44$\arcsec. The resulting superposition of the MIR map and the radio sources is displayed in Fig.~\ref{NGC253_timmi}b. We do not believe this blind registration to be the correct one, since the MIR string and the radio string are shifted one with respect to the other. However, we note that under this registration, M4 is coincident with the radio nucleus. 

(b) We try to align the MIR and radio peaks. To do so, we simply minimise the sum, over all the MIR sources, of the distances to the closest radio source. Fig.~\ref{NGC253_timmi}c shows the resulting map. This registration suggests, this time, that source M3 is the counterpart of the radio nucleus. The superposition is globally much more convincing than the blind one. In this case, the B1950 coordinates of M1 can be derived:  $\alpha$=00$^{\rm h}$45$^{\rm m}$5.703$^{\rm s}$ \& $\delta$=-25\degr33\arcmin40.57\arcsec. These coordinates are not consistent, within the quoted error boxes, with the absolute coordinates of M1 derived by \citet{Keto93}, but are consistent with those, with larger error bar, by \citet{Pina92} and \citet{Kalas94}.  

Despite the inconsistency with Keto's measurement, we favour the second registration, for which the string of radio sources is well aligned with the MIR emission. In none of these registrations does source M1 have a radio counterpart. The lack of radio emission from M1, already discussed by \citet{Forbes00}, is confirmed by our study and remains mysterious. And finally, our preferred registration places the radio nucleus in coincidence with the MIR source M3.

\subsection{Circinus: still point like at 0.6\arcsec~resolution}
\label{Circinus}
\begin{figure}[htbp]
\begin{center}
\resizebox{8cm}{!}{\includegraphics*[scale=1.]{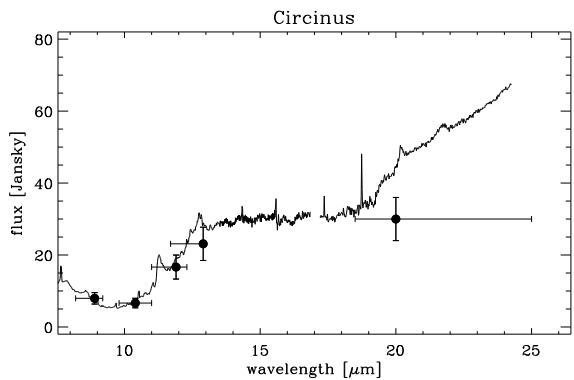}}
\caption{Photometric measurements for the Circinus galaxy. The filled circles show our measurements. The ISO spectrum from \citet{Sturm00} is overplotted for comparison.}
\label{sed Circinus}
\end{center}
\end{figure} 
\begin{figure}[htbp]
\begin{center}
\resizebox{7cm}{!}{\includegraphics*[scale=1.]{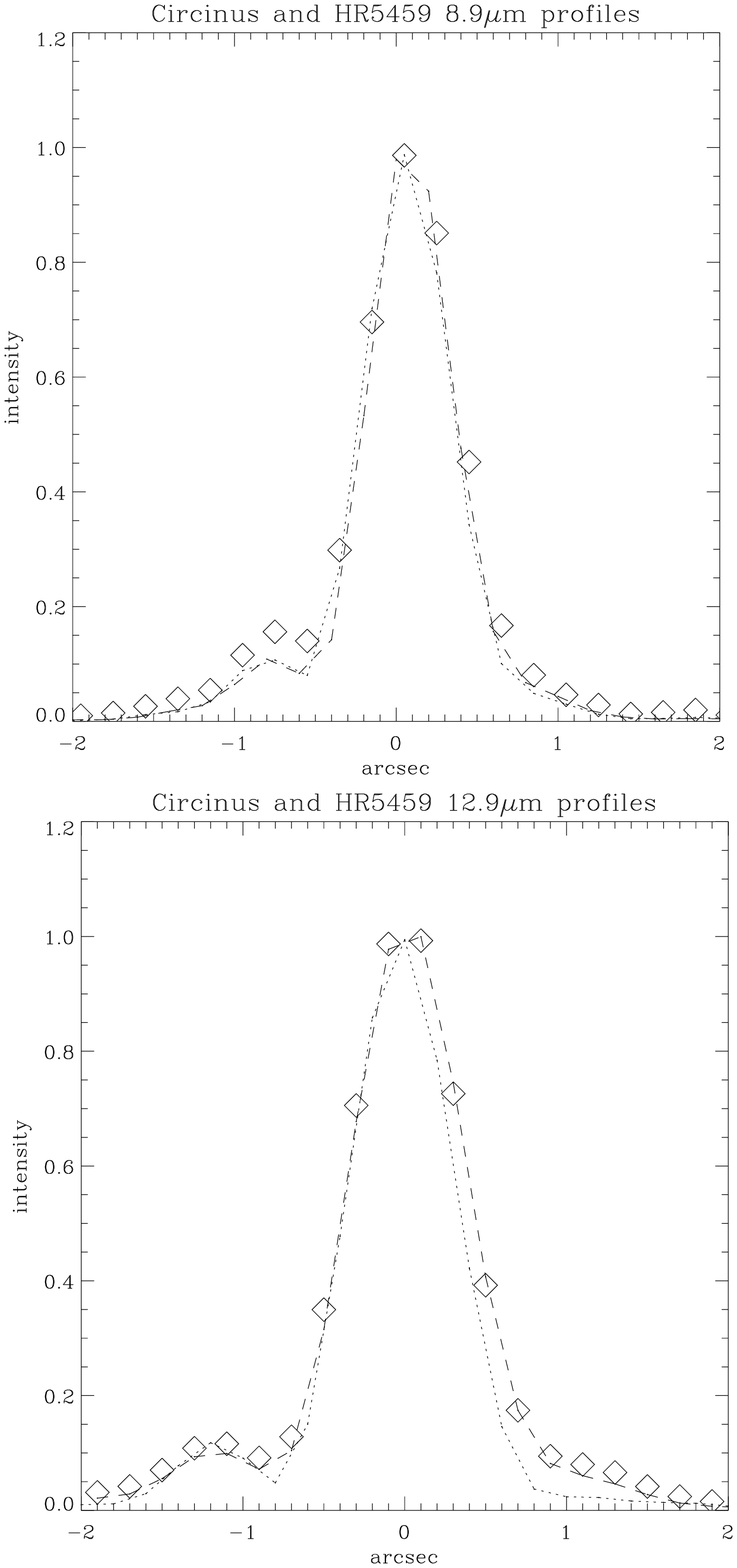}}
\caption{Circinus: profiles along the E-W direction at 8.9\micro~(top) and 12.9\micro~(bottom). The squares feature the profile of Circinus and its sampling, while the dashed and dotted lines show the profiles of the standard star HR5459, acquired before and after Circinus, respectively.}
\label{Circinus profile}
\end{center}
\end{figure} 

Circinus is the closest type 2 AGN, at a distance of only 4\,Mpc. The nuclear region harbours a heavily obscured AGN \citep[$N_H \geq 10^{24} \mbox{cm}^{-2}$ based on \Xray~data by][]{Matt99}. It also exhibits a prominent ionisation cone extending to the NW, and a starburst ring at a radius of about 10\arcsec~($\sim$100\,pc) from the nucleus \citep{Davies98}. 
Circinus has been previously imaged at high resolution with TIMMI \citep{Siebenmorgen97} and MANIAC \citep{Krabbe01}, both showing marginal extension of the central source. The latter study also revealed the presence of a low surface brightness diffuse halo of radius $\sim$ 10\arcsec. 

Aperture photometry was performed to obtain the N-band fluxes for Circinus within an aperture of 2\arcsec~radius. This radius was optimised using a curve of growth analysis. The background was measured in an annulus around this aperture (with inner and outer radii 2\arcsec~and 4\arcsec, respectively). The \Tim~data points are plotted in Fig.~\ref{sed Circinus}. The 14\arcsec$\times$20\arcsec~aperture ISO-SWS spectrum by \citet{Sturm00} is shown.The N-band \Tim~measurements are in good agreement with the ISO spectrum, showing that the entire MIR emission comes from the nuclear source. At 20\micro, the discrepancy between the \Tim~data and the ISO spectrum is difficult to interpret in a decisive manner. Observing and calibrating ground based data in this window is a difficult task. Moreover, the 20\micro~emission is expected to be more extended than the 10\micro~one, since it traces cooler material. Undetected diffuse emission on a scale of the order of 10\arcsec, could be responsible for this difference.

On the \Tim~data, the comparison between the AGN and the corresponding standard star profiles does not reveal any sign of resolved emission in Circinus on a scale 0.3-0.4\arcsec~(5.5-7.5\,pc). The \Tim~image reaches a 1$\sigma$ background level of 25\,mJy\,arcsec$^{-2}$, which is insufficient to detect the broad N-band halo emission claimed by Krabbe et al. in 2001 (17\,mJy\,arcsec$^{-2}$).

\subsection{NGC1808: MIR detection of radio sources}
\label{sect:NGC1808}
\begin{figure*}[htbp]
\begin{center}
\resizebox{18cm}{!}{\includegraphics*[scale=1.]{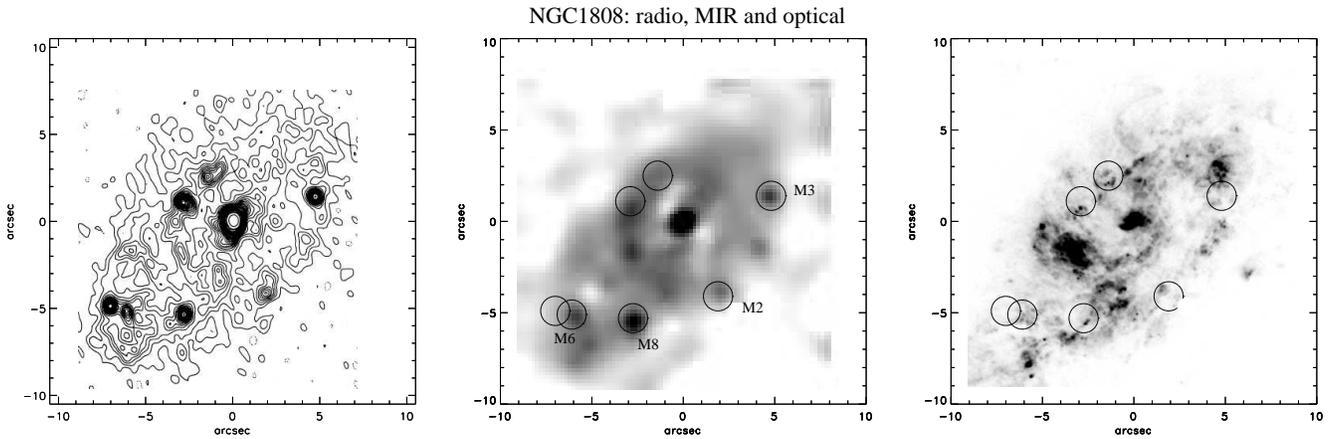}}
\caption{NGC1808: from left to right, radio, MIR and optical images. The map at 3.6\,cm is from \citet{Collison94}; our \Tim~map at 12.9\micro~is displayed with a non-linear flux scale in order to highlight faint structures; the HST image (F658N) is from the HST archive. On the MIR and optical maps, the locations of the main radio sources are marked by circles of radius 0.8\arcsec.}
\label{NGC1808}
\end{center}
\end{figure*} 
NGC1808 is a morphologically peculiar spiral galaxy (Sbc pec) located at a distance of 10.9\,Mpc \citep{Koribalski96}. The nature of its nuclear source is not yet clearly established. \citet{Veron-Cetty85} concluded to the presence of an AGN, while \citet{Forbes92} proposed instead that the nucleus consists of a SNR embedded within a metal-rich HII~region. NGC1808 has been detected in hard \Xrays~\citep{Awaki93,Junkes95}, suggesting that an embedded AGN is indeed present in the nuclear region. However, \citet{Krabbe01} show that, in a MIR vs. \Xray~flux diagram, the location of NGC1808 argues against the presence of an AGN.

The central region (inner $\sim$750\,pc) of \object{NGC1808} is undergoing an episode of enhanced star formation. There were early reports on the presence of ``hot spots'' in its complex nuclear region by \citet{Morgan58} and \citet{Sersic65}. Strong PAH emission was observed by \citet{Siebenmorgen01} from the star-forming region. Radio continuum observations \citep{Saikia90,Collison94} revealed a population of compact radio sources, most of which do not coincide with the optical hot spots. According to \citet{Saikia90}, they are more likely related to young SNRs than due to the thermal emission from giant HII~regions. A high resolution K-band image was published by \citet{Tacconi-Garman96}, showing the presence of many distinct and compact ($\leq 0.7\arcsec$) sources spread all over the circumnuclear region. These authors argue that such sources are likely to be clusters containing hundreds of massive red super-giants. Not all of the radio emitting sources have counterparts in the map by \citet{Tacconi-Garman96}. Particularly, the three brightest circumnuclear radio sources have no NIR counterpart. 

\citet{Krabbe01} found a nuclear flux of 0.33\,Jy through an N-band (8-13\micro) filter. This value is in reasonable agreement with ours (see Table~\ref{results}) although a precise comparison is not possible because of the different filters used. The 12.9\micro~map of the central region of NGC1808 is reproduced in Fig.~\ref{NGC1808} together with the 3.6\,cm image obtained by \citet{Collison94} and an HST archive image in the optical (F658N). The radio and MIR maps are strikingly similar. The most intense radio sources are featured by circles of arbitrary 0.8\arcsec~radius on the MIR and optical maps. Several of the intense radio sources have a counterpart on the MIR image. Particularly, this comparison shows that M2, M3, M6 and M8 are also radio sources. On the HST image, counterparts for the MIR/radio sources are not clearly detected,  but may be identified to faint sources that can be seen inside the circles in Fig.~\ref{NGC1808}. M8 is the source that has most likely an optical counterpart. A comparison between a higher resolution MIR map and the HST image is needed in order to investigate which optical sources are the counterparts of the MIR/radio sources. Still, it is clear that the regions of most intense optical emission (the hot spots) do not coincide with either radio or MIR sources. The nature of these MIR/radio sources will be discussed in detail in Sect.~\ref{discussion2}.

\subsection{NGC1365: MIR detection of radio sources}
\label{sect:NGC1365}
\begin{figure*}[htbp]
\begin{center}
\resizebox{18cm}{!}{\includegraphics*[scale=1.]{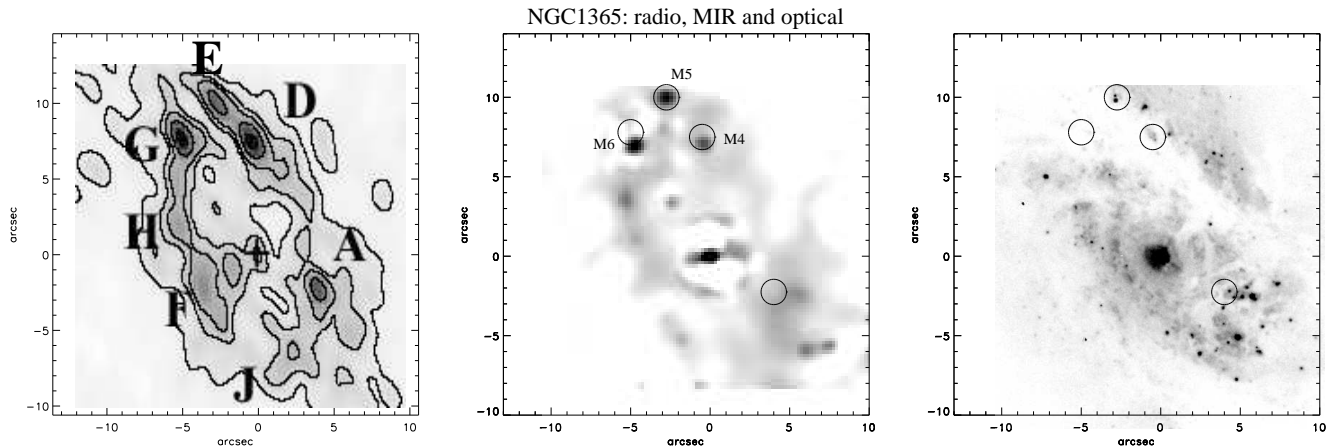}}
\caption{NGC1365: from left to right: radio, MIR and optical images. The map at 3\,cm is from \citet{Forbes98}; our \Tim~map at 11.9\,\micro~is displayed with a non-linear flux scale in order to highlight faint structures and the HST image (F814W) is from the archive. On the MIR and optical maps, the locations of the main 3\,cm sources are marked by circles of radius 0.8\arcsec.}
\label{NGC1365}
\end{center}
\end{figure*} 

A global picture of NGC1365 can be found in the dedicated review by \citet{Lindblad99}. This spiral galaxy with a prominent bar has been classified as SBb(s)I \citep{Sandage81}. An estimate of its distance is discussed in \citet{Lindblad99} and is currently set to 18.6$\pm$0.6\,Mpc. At this distance, 1\arcsec~corresponds to 90\,pc. The nuclear region harbours an AGN witnessed by the presence of both broad and narrow Balmer emission lines, as well as narrow and strong forbidden emission lines of high excitation \citep{Veron80,Alloin81}.  It is classified as a type 1.5 AGN, but varies from type 1 to type 2 according to the level of its hard radiation emission.

It has long been known to show interesting structures in its nuclear region \citep{Sersic65}. In the optical, this region shows several bright hot spots within 1\,kpc around the AGN. The HST images resolve the circumnuclear region into a large number of bright, compact star clusters of only a few parsec radii \citep{Kristen97,Lindblad99}. In the radio domain, continuum observations have been performed at 20, 6 and 2\,cm with the VLA \citep{Sandqvist82,Saikia94,Sandqvist95} and at 6 and 3\,cm by \citet{Forbes98} with the Australian Compact Array (ATCA). These maps show that the nucleus is surrounded by several non-thermal radio sources, while the AGN itself is not a particularly strong radio source. The radio sources do not coincide with the bright optical knots. A set of NIR images of the nuclear region of NGC1365 at very good spatial resolution ($\sim$0.3\arcsec) were taken during the commissioning of ISAAC at the VLT by \citet{Moorwood98}: very red (K-L $\approx4$) counterparts to the brightest radio sources were detected, suggesting the presence of deeply embedded sources.

Very few MIR observations of NGC1365 have been performed to date. \citet{Frogel82a} made a series of MIR measurements that are only consistent within a factor of two with our recent measurements. Such a difference can have various origins (differences in filter shape, aperture, and even, over this 20 years period, genuine variations of the MIR emission of the AGN). The map at 11.9\micro~reveals a 4\arcsec~extension centred on the nucleus (M1) along the E-W direction~(Fig.~\ref{All contours}). This extension does not clearly appear on the other maps. At 10.4, 11.9 and 12.9\micro, the nuclear source (M1) itself presents a slight (2\arcsec) extension in the E-W direction which is absent in the other MIR sources in the field. Therefore, we believe that this extension is real. Indeed, to the East of the nucleus, a radio ``jet-like'' structure between the nucleus and the radio source F has been observed by \citet{Sandqvist95}. Thus the MIR and radio emissions of this extended feature might be related. Figure~\ref{NGC1365} compares, at the same scale, the maps in the visible, in the radio and in the MIR. Both sources M2 and M3, are clearly detected at 10.4\,\micro, while at 11.9\,\micro~only M3 shows up and neither of them is detected at 12.9\,\micro. This indicates that M2 and M3 are not strong dust emitters, with a rather flat or decreasing MIR spectrum and no PAH emission. 
The comparison between the radio and the MIR maps shows that, as in the case of NGC1808, the bright radio sources have a counterpart in the MIR, while they do not coincide with any strong optical feature. The coincidence between M6 and the radio source G is not perfect, if using the map by \citet{Forbes98}. Comparison with the 2\,cm map by \citet{Saikia94} gives a much better coincidence. This kind of discrepancies may come from the uncertainty in the reconstruction of radio maps when the emission is from an extended component together with point-like sources. We conclude that sources M4, M5 and M6 are the respective counterparts of radio sources D, E and G. The optical HST image clearly shows the presence of a dust lane to the north of the AGN. At the location of M6, no optical source is detected, which is understandable as M6 lies in the middle of the dust lane. Two optical sources are present at the location of M5 and might be its counterpart. An extended cone-shaped optical source appears at the location of M4, and may be related to it. The nature of the MIR/radio sources in NGC1365 is discussed, together with those observed in NGC1808, in Sect.~\ref{discussion2}.

\subsection{NGC7469}
\label{NGC7469}

NGC7469 exhibits a type 1 AGN hosted in an SBa galaxy. A significant fraction of the nuclear luminosity originates in a circumnuclear ring of star formation (1\,kpc diameter), which is seen face-on and thus appears more or less symmetrical around the AGN. The circumnuclear ring has been detected in the optical \citep[speckle imaging by][]{Mauder94}, NIR \citep{Marco98} and MIR \citep{Miles94}, while PAH emission has been observed by \citet{Mazzarella94} and \citet{Miles94}. Non-thermal radio continuum emission is detected both as an extended component and as several knots located 1.5\arcsec~to 2\arcsec~away from the nucleus \citep{Ulvestad81,Condon82,Wilson91}, hence belonging to the star formation ring. 

The 10.4\micro~flux measured from the \Tim~dataset for the central source is in good agreement with the N-band measurement from \citet{Gorjian04}. The morphology seen on the \Tim~map at 11.9\micro~is consistent with high resolution NIR images, particularly the 0.4\arcsec~resolution J and H SHARP maps published by \citet{Genzel95}: the circumnuclear ring exhibits enhanced brightness to the NE and SW. Surprisingly, the 10.4\micro~image has a quite different aspect and reveals only an extension along the $\sim$N-S direction. A simple interpretation is that, except along the N-S axis, the extended MIR emission is dominated by PDR emission, with low continuum at 10.4\micro~and intense PAH emission in the 11.9\micro~filter. An additional N-S micro-bar with large dust content could explain the observed intense N-S emission at 10.4\micro.

\subsection{Unresolved galaxies}
\label{unresolved galaxies}
\object{NGC2992} is an almost edge-on Sa galaxy, interacting with \object{NGC2993}. It is at a distance of 30\,Mpc, which gives a scale of 150\,pc/\arcsec. It harbours a type 2 AGN, located behind a prominent dust lane extending along the major axis of the galaxy. The radio source at the location of the AGN is peculiar in shape (shape of a 'figure 8') and extends up to 450\,pc \citep{chapman00}. The nuclear region of NGC2992 was observed through the \Tim~10.4 and 12.9\micro~filters. We did not detect any conspicuous diffuse/extended emission down to a level of 35\,mJy\,arcsec$^{-2}$ ($1\sigma$) at 12.9\micro. The AGN appears as a point-like source with FWHM=0.9\arcsec~on the 10.4\micro~image. The flux measurements are in agreement with the measurements by \citet{Gorjian04}. The 12.9/10.4 flux ratio has a value of 2.44, compatible with both the AGN and PDR templates, and thus PAH bands may be present. A detection of the 11.3\micro~PAH band was claimed by \citet{Roche91}, although \citet{Mizutani94} did not confirm this result as they did not find the PAH signature at 3.3\micro.

\object{NGC5995} hosts a type 2 AGN and is located at 100\,Mpc, leading to a scale of 490\,pc/\arcsec. It has been observed with \Tim~through the 8.9 and 11.9\micro~filters. Comparable flux values were found in both filters, suggesting emission from a PDR. The 11.9\micro~flux is found to be slightly lower than, but consistent with, the IRAS flux value, implying that most of the MIR emission is from the central source.  

IZw1 is the archetypal narrow-line type 1 AGN.
It is considered as the closest QSO because of its large intrinsic optical luminosity. At a distance of 245\,Mpc, the linear scale is 1200\,pc/\arcsec. \citet{Schinnerer98} find a strong starburst in a circumnuclear ring of $\sim$1.5\arcsec~in diameter (1.8\,kpc). 
IZw1 was observed with \Tim~at 10.4, 11.9 and 12.9\micro. The fluxes are found to be in good agreement with the IRAS ones, indicating that the bulk of the MIR emission comes from the unresolved nucleus. Interestingly, the 12.9\micro~flux is significantly lower than the 11.9\micro~flux (as in the case of PDR emission), and is at the same level as the 10.4\micro, which is not expected for a PDR emission.  This suggests the presence of the silicate feature in emission. Spectroscopy in the MIR should allow to confirm this prediction.

\object{IIZw136} is a type 1 AGN, 252\,Mpc away (scale of 1200\,pc\arcsec). It has been observed through the 10.4 and 11.9\micro~filters: equal flux values are found in both filters. Such a flux ratio, 1$\pm$0.3, is consistent with either the HII~region template or the AGN template if the 9.7\micro~silicate feature is not as deep as in the template.

\section{Active galactic nuclei: building the SED}
\label{discussion1}

\begin{table*}[htbp]

\caption[]{Compiled NIR-MIR photometry for the galaxies in our sample. The MIR data presented here are the integrated fluxes over the aperture given in the table.}

\begin{center}

\begin{tabular}{|ccccccc|} \hline

Name	 	&Wavelength 	& Aperture		& Projected	&	Flux 			& Reference 		& $\rm log(\nu L_{\nu})$ 	\\

  		&[\micro]	& [arcsec] 		& size [pc]	&	[mJy] 			& 	    		& [$\rm erg.s^{-1}$]  		\\ \hline

Circinus 	& 1.265         & 0.19			& 3.6		&	1.6$\pm	   0.2$	&\citet{Prieto04}	& 39.9				\\ 
2.5\,Mpc	& 1.66		& 0.19			& 3.6		&	4.77$\pm   0.5$      &\arcsec		& 40.2      			\\
		& 2.18		& 0.19			& 3.6		&	19$\pm	    1.9$     	&\arcsec		& 40.7      			\\
		& 3.80		& 0.19			& 3.6		&	380$\pm	    38$      &\arcsec		& 41.8      			\\
		& 4.78		& 0.19			& 3.6		&	1900$\pm    190$     	&\arcsec		& 42.4      			\\
		& 8.9		& 2			& 38		&	7950$\pm    1590$     	& This work		& 42.7      			\\
		& 10.4		& 2			& 38		&	6650$\pm    1330$     	&\arcsec 		& 42.6      			\\
		& 11.9		& 2			& 38		&	16650$\pm   3330$     	& \arcsec		& 42.9      			\\
		& 12.9		& 2			& 38		&	23400$\pm   4680$     	& \arcsec		& 43.0      			\\
		& 20		& 2			& 38		&	30000$\pm   6000$     	& \arcsec      		& 42.9		      		\\ \hline
NGC1808		& 2.2		& 2			& 106		&	43.5$\pm    4.4$     	&\citet{Galliano05a}     & 41.9  			\\
10.9\,Mpc	& 3.5		& 2			& 106		&	36.9$\pm    7.4$     	&\arcsec	        & 41.7  			\\	
		& 4.8		& 2			& 106		&	19.1$\pm    3.8$     	&\arcsec		& 41.2				\\
		& 10.4		& 2			& 106		&	255$\pm	    51$     	&This work		& 42.0      			\\			   
		& 11.9		& 2			& 106		&	620$\pm	    124$     	&\arcsec		& 42.3      			\\			   
		& 12.9		& 2			& 106		&	970$\pm	    194$      &\arcsec		& 42.5      			\\ \hline		   
NGC1365		& 2.2		& 2			& 106		&	78$\pm	    7.8$     	&\citet{Galliano05a}     & 42.6  			\\			   
18.6\,Mpc	& 3.5		& 2			& 106		&	205$\pm	    41$   	  &\arcsec 	        & 42.9  			\\			   
		& 4.8		& 2			& 106		&	177$\pm	    35$     &\arcsec		& 42.7				\\			   
	       	& 8.9		& 2			& 180		&	410$\pm	    82$     &This work		& 42.8      			\\			   
		& 10.4		& 2			& 180		&	440$\pm	    88$     &\arcsec		& 42.7      			\\			   
		& 11.9		& 2			& 180		&	510$\pm	    102$     &\arcsec		& 42.7      			\\			   
		& 12.9		& 2			& 180		&	110$\pm	    220$     &\arcsec		& 43.0      			\\ \hline		   
NGC2992		& 2.18		& 3			& 450		&$^a$	2.8$\pm	   0.2$     &\citet{Alonso-Herrero01}&41.6      			\\			   
301.\,Mpc	& 3.8		& 3			& 450		&$^a$	22.7$\pm    2.0$     	&\arcsec		& 42.3  			\\		   
		& 4.8		& 3 			& 450		&$^a$	35.7$\pm    10.7$     	&\arcsec		& 42.4  			\\			   
		& 10.4		& 2			& 300		&      230$\pm	    46$     	&This work		& 42.9      			\\			   
		& 12.9		& 2			& 300		&      565$\pm	    110$     	&\arcsec		& 43.2      			\\ \hline		   
NGC7469		& 1.1		& 3			& 960		&$^a$	16.2$\pm    2.4$     	&\citet{Alonso-Herrero01}&43.4				\\			   
66\,Mpc		& 1.6		& 3			& 960		&$^a$	39$\pm	    5.9$     	&\arcsec		& 43.6      			\\			   
		& 2.22		& 3			& 960		&$^a$	67.8$\pm    17.0$     	&\arcsec		& 43.7      			\\			   
		& 3.80		& 3			& 960		&$^a$	159$\pm	    4.8$     &\citet{Ward87}		& 43.8      			\\			   
		& 4.78		& 3			& 960		&$^a$	259$\pm	    31$     &\arcsec		& 43.9      			\\			   
		& 10.4		& 2			& 640		&	310$\pm	    62$     &This work		& 43.7      			\\			   
		& 11.9		& 2			& 640		&	565$\pm	    113$     &\arcsec		& 43.9      			\\ \hline		   
NGC5995		& 2.2		& 10			& 3200		&	38$\pm	   0.8$     	&\citet{Hunt99}		& 43.8				\\ 			   
		& 8.9		& 2			& 980		&	310$\pm	    62$     &This work		& 44.1      			\\			   
100\,Mpc	& 11.9		& 2			& 980		&	300$\pm	    60$     &\arcsec		& 44.0      			\\ \hline		   
IZw1		& 3.5		& 8.5			& 10000		&	110$\pm	    11$     &\citet{Rieke78} 	& 44.8      			\\			   
245\,Mpc	& 10.4		& 2			& 2370		&	375$\pm	    75$     &This work		& 44.9      			\\			   
		& 11.9		& 2			& 2370		&	425$\pm	    85$     &\arcsec		& 44.9      			\\ 			   
		& 12.9		& 2			& 2370		&	325$\pm	    65$     &\arcsec		& 44.7				\\ \hline		   
IIZw136		& 10.4		& 2			& 2440		&	130$\pm	    26$     &This work		& 44.5      			\\ 			   
252\,Mpc	& 11.9		& 2			& 2440		&	130$\pm	    26$     &\arcsec		& 44.4      			\\ \hline                  
														  

\end{tabular}

\end{center}
$^a$ Non-stellar contribution, as computed by \citet{Alonso-Herrero01}

\label{tab:SED} 

\end{table*}

\begin{figure*}[htbp]
\begin{center}
\resizebox{18cm}{!}{\includegraphics*[scale=1.]{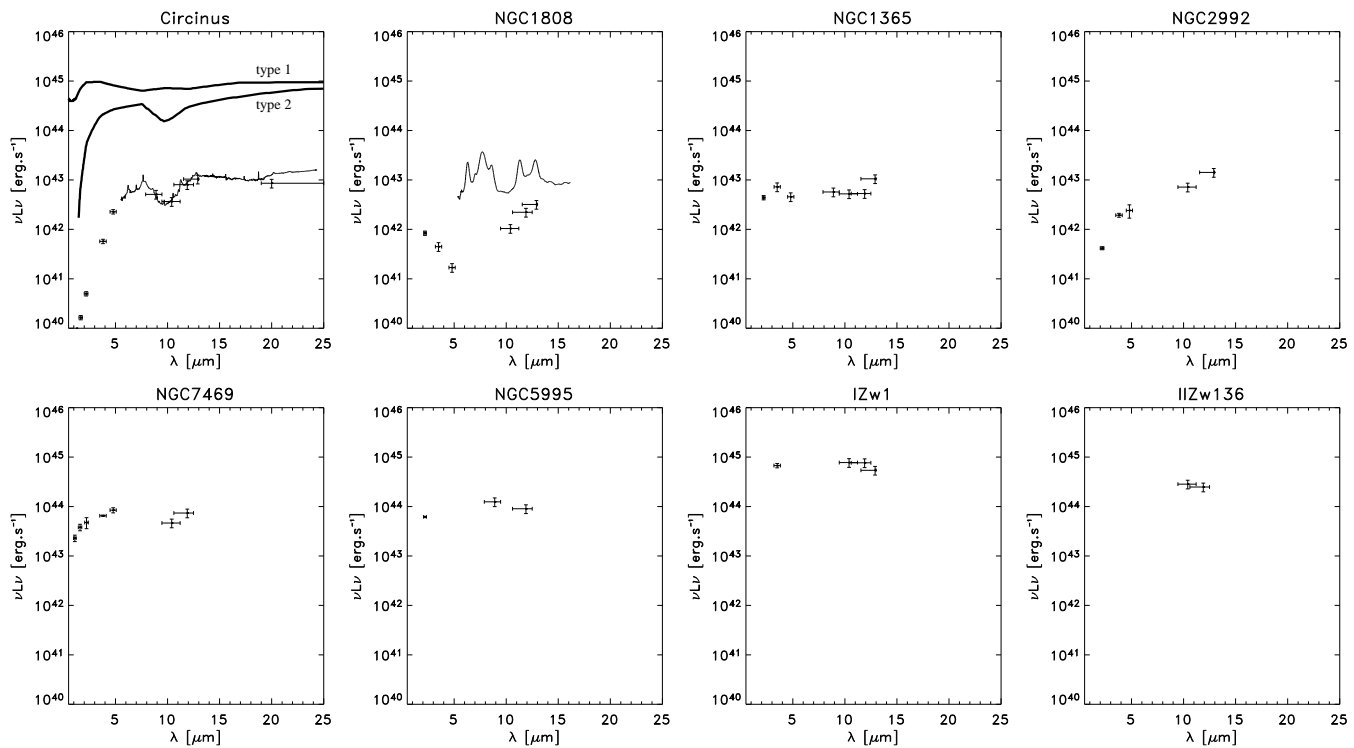}}
\caption{NIR-MIR spectral energy distributions for the AGN in our sample. Intrinsic luminosities have been calculated using the distances reported in Table~\ref{obs_summary} and they are provided in Table~\ref{tab:SED}. The vertical bar attached to each point represents the photometric uncertainty, while the horizontal bar represents the width of the filter through which the measurement has been performed. Large aperture ISO-SWS spectra (thin continuous line) are overplotted for Circinus and NGC1808. On the top-left panel, we have overplotted with thick lines modelled typical SED for type\,1 and type\,2 AGN \citep[][ Fig.\,1, second model]{Granato97}. For the sake of clarity, the flux level of these model SED has been set at a large value. }
\label{fig:SED}
\end{center}
\end{figure*} 

This short section focuses on the SED of the AGN in our sample. Historically, diagnostics of AGN SED were preferably applied to high luminosity objects, in order to avoid complication due to the mixing of light from the AGN and from the host galaxy. Yet, the study of weaker close-by AGN is important since their proximity allows their inner structure to be unveiled down to physical scales of a few parsecs, corresponding to the expected sizes of some of their components. In such cases, the main difficulty arises from the fact that the contrast between the AGN and its host galaxy is low, particularly when the AGN is surrounded by regions of starburst activity. Then, large aperture photometric measurements can greatly differ from the actual AGN fluxes.

As the observational constraints on the torus models \citep{Krolik86,Pier93,Granato94,Efstathiou95,Granato97} come from the shape of the SED between 1 and 20\micro, and from the depth of the 9.7\micro~silicate feature, it is particularly important to isolate the AGN and measure its flux free from contamination by other components. We provide a compilation of small aperture measurements of the NIR-MIR fluxes for our sample AGN (Table~\ref{tab:SED}). The corresponding SED are shown in Fig.~\ref{fig:SED}. On the top-left panel, along with the SED of Circinus, typical SED models have been overplotted for type\,1 and type\,2 AGN, as predicted using the radiative transfer code by \citet{Granato97}. Type 1 AGN show a flat SED (in $\nu L_{\nu}$), while type 2 AGN show a SED with a steep fall off at NIR wavelengths. 

\begin{itemize}

\item For Circinus, the compiled SED is from high resolution NACO NIR data and our MIR measurements. The SED is fully consistent with the ISO spectrum (overplotted) and shows that the bulk of the NIR-MIR emission comes from a type 2 AGN. 

\item For NGC1808, we plotted new NIR small aperture measurements \citep{Galliano05a} together with the \Tim~data. The MIR points are distributed similarly to those of Circinus, although slightly shifted towards lower luminosities. This supports the idea that the nucleus of NGC1808 is indeed an AGN. However, at shorter wavelengths, the measurements show a SED increasing towards short wavelengths, suggesting the presence of a bright, stellar core. The overplotted large aperture ISO spectrum is from \citet{Siebenmorgen01}. Its flux level, about 10 times larger than that of the AGN, demonstrates that the central 1\,kpc region is dominated by the emission from the starburst.

\item The AGN in NGC1365 is thought to be a type 2. Hence, its SED is expected to be decreasing from the MIR to the NIR. New small-aperture NIR measurements \citep{Galliano05a} are plotted together with the \Tim~fluxes and show that, in contrary to expectations, the SED is flat and consistent with that of a type\,1 AGN. The presence of a local maximum at 3.5\micro~suggests a contribution from PAH. The AGN in NGC1365 is thus a candidate for the detection of PAH around AGN.

\item The NIR-MIR SED of NGC2992 and NGC7469 nuclei are respectively consistent with type 2 (steep SED) and type 1(flat SED) AGN. 

\end{itemize}       

The remaining sources shown in Fig.~\ref{fig:SED} (NGC5995, IZw1 and IIZw136) are described in Sect.~\ref{unresolved galaxies}. Detailed and unambiguous modelling of the SED of these AGN requires the use of additional data: as shown in \citet{Galliano03b} photometric measurements alone allow the fit of many different models of tori. A way to lift this degeneracy would be to use spatially resolved information, which means to observe these AGN with IR interferometers.

\section{Circum-AGN environments: embedded star clusters in NGC1808 and NGC1365}
\label{discussion2}
\begin{figure*}[htbp]
\begin{center}
\resizebox{15cm}{!}{\includegraphics*[scale=1.]{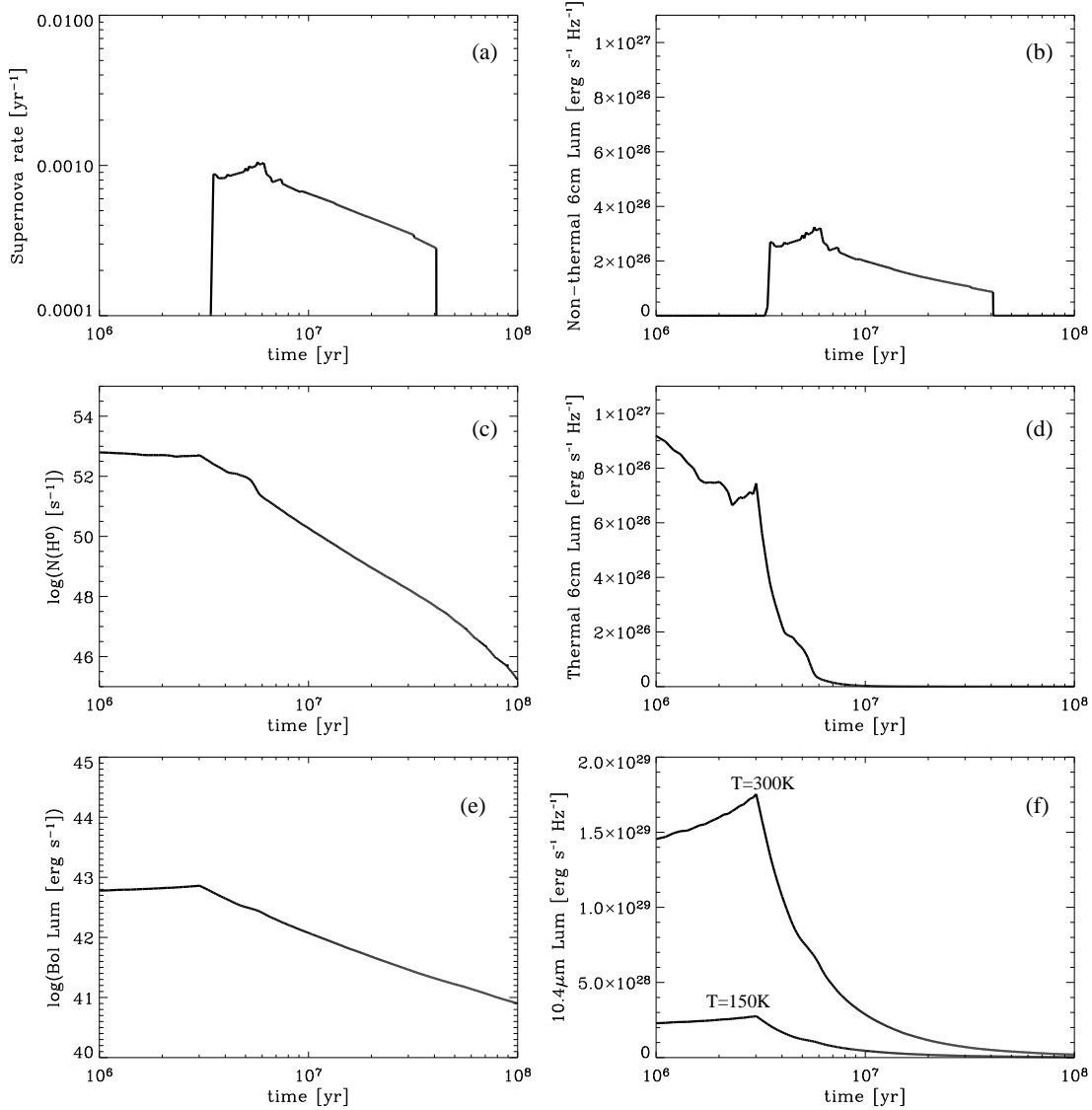}}
\caption{Evolution of the properties of a star cluster. We consider a Starburst99 model for a $10^6$ \msol~cluster with instantaneous star formation and Salpeter IMF. On the plots to the left, evolution of the star cluster intrinsic properties are shown. On the plots to the right, predicted observational properties as a function of time are shown. An index $\alpha=-0.9$ is used for the SNR radio emission. (a) Rate of supernova events per year. (b) Non-thermal luminosity at 6\,cm produced by the SNR. (c) Lyman photon production rate. (d) Thermal luminosity at 6\,cm produced in the HII regions. (d) Bolometric luminosity of the cluster. (e) Evolution of the 10.4\,\micro~luminosity if all the energy of the stars is reprocessed in a black-body at respectively 300\,K and 150\,K.}
\label{model}
\end{center}
\end{figure*} 

MIR imaging at high spatial resolution provides the opportunity of detecting young massive dust embedded clusters in starbursting galaxies. These clusters emit a bright IR continuum, infrared nebular emission lines and are strong cm radio sources. The IR continuum arises from the dust heated by stars in the cluster, while the lines and usually most of the cm continuum are produced by the HII~regions surrounding the massive stars in the cluster and/or by supernova remnants (SNR). 

To date, only a very limited number of such embedded clusters are known, namely the clusters in He2-10 \citep{Kobulnicky99,Beck01,Vacca02}, NGC5253 \citep{Turner98,Gorjian01,Vanzi04}, the Antennae \citep{Mirabel98,Gilbert00}, IIZw40 \citep{Beck02} and SBS0335-052 \citep{Plante02}. The embedded clusters are found to be young (a few Myr), massive ($\sim 10^6$\,\msol)~and highly obscured ($\rm A_V=$5 to 100). 

The fact that the bright compact radio sources surrounding the AGN in NGC1808 and NGC1365 are found to be also intense MIR sources suggests that they are related to very dusty objects. Since these sources have no comparably intense optical counterpart (except NGC1365/M5), they are very likely to be embedded star clusters. Still, this interpretation needs quantitative support: in the following, we present comparisons between  the predictions from a simple model for embedded star clusters and our observations in the MIR and the radio. We deliberately chose to consider a simple model, given the still limited amount of data available for comparison. The calculations in this section only intend to give support to the interpretation in terms of embedded clusters and to give rough estimates of their properties (order of magnitude for their age and mass). Detailed and more realistic modelling of such objects is postponed until new data, especially spectroscopic measurements, will become available. Throughout, the radio spectral indexes are noted $\alpha$, with $S_{\nu} \propto {\nu}^{\alpha}$.
\subsection{Emission from young star clusters}
\label{emission from young star clusters}
Prior to being identified in the MIR, the radio sources surrounding the nuclei in NGC1808 and NGC1365 were commonly interpreted as radio supernovae or as supernova remnants (SNR). This was suggested by the negative values of their cm spectral indexes, indicating non-thermal emission (the radio measurements are given in Table~\ref{table result}). However, the interpretation in terms of \textbf{isolated} (i.e. not associated with a star cluster) radio supernova or SNR  can now be discarded since these objects would not show up as IR sources for more than a few years \citep{Weiler86} and therefore would have a low probability of being detected as such. As we do detect them in the MIR, we consider instead that they are radio supernova or SNR \textbf{associated with a young star cluster}, where their probability of occurrence is large.

The observed radio spectral index for the sources range from steep to rather flat, but always indicate an important share of non-thermal processes. Two types of radio emission are expected from young compact clusters: free-free thermal emission from the HII regions surrounding the massive stars ($\alpha \sim -0.1$ where $\alpha$ can become positive in the case of optically thick emission), and non-thermal emission from both radio supernovae and SNR ($\alpha \sim -0.9$). We examine hereafter the three possibilities. For simplification, we only consider optically thin emission, being conscious that such an assumption might not always hold true.  

The first mechanism for the production of radio continuum is free-free emission in the HII regions surrounding massive stars. The observed flux depends on the rate of emission of H ionising photons ($E \geq 13.6 \rm eV$). Following \citet{Turner98}, in the case of optically thin radio emission, the cm flux is related to the number of ionising photons by the following relation: 
\begin{equation}
\rm
\left(\frac{S_{\nu}}{mJy}\right)_{HII}=1.45\,10^{-50} \left(\frac{d}{Mpc}\right)^{-2} \left(\frac{\nu}{GHz}\right)^{-0.1} \left(\frac{\dot{n}_{N}}{s^{-1}}\right)
\label{thermal}
\end{equation}
where $\rm S_{\nu}$ is the flux density at frequency $\nu$, $\rm \dot{n}_{N}$ is the rate of emission of ionising photons, and $\rm d$ is the distance to the source. 

Some synchrotron sources appear and disappear within periods of months to years following Type II or Type Ib supernovae events and can be very intense. However, because of their short lifetime, and because we observe simultaneously several sources in NGC1365 and NGC1808, these radio supernovae are unlikely to be the dominant contributors. This contribution is therefore discarded in the following discussion.

Finally, the radio emission from SNR starts about 50 years after the supernova event and persists for hundreds to thousands of years. We adopt the simple recipe given by \citet{Condon90}, relating the non-thermal radio emission from galaxies to the type II supernova rate: 

\begin{equation}
\rm
\left(\frac{S_{\nu}}{mJy}\right)_{SNR}=10^6\left(\frac{d}{Mpc}\right)^{-2} \left(\frac{\nu}{GHz}\right)^{\alpha} \left(\frac{\dot{n}_{SN}}{yr^{-1}}\right)
\label{non thermal}
\end{equation}

where $\rm S_{\nu}$ is the flux density at frequency $\nu$, $\alpha$ is the spectral index, d is the distance of the object and $\dot{n}_{SN}$ is the supernova rate in yr$^{-1}$. This mechanism is bound to play a role and will be considered in the discussion.     



Let us evaluate the relative share of the two contributing mechanisms, free-free emission and non-thermal emission from SNR, along the evolution of a star cluster. As argued previously, the observed cm spectral indexes indicate that the emission has an important share of non-thermal emission produced by SNR. Using the stellar population evolutionary model Starburst99 developed by \citet{Leitherer99}, we can compute, for a given cluster, the time evolution of the supernova rate and of the emission rate of ionising photons. We choose a stellar population model with instantaneous star formation, solar metallicity, a total mass of $10^6$\,\msol~and a Salpeter initial mass function. 

Instantaneous star formation is a straightforward assumption to model the star cluster evolution. Is solar metallicity a reasonable assumption? In spiral galaxies, heavy elements are often found to be overabundant -with respect to solar- in the central region of the galaxy and also to exhibit an abundance gradient across the galaxy disc. In the case of NGC1365, the oxygen abundance gradient  (derived from the analysis of HII regions) is rather mild (0.02 dex/kpc) and gives an extrapolated oxygen abundance at the centre: 12+log(O/H)= 9.1 \citep{Alloin81}. With a solar oxygen abundance of 8.66 in the same unit \citep{Asplund04}, the metallicity at the centre of NGC1365 is found to be around 3 times solar. The Starburst99 models presented in \citet{Leitherer99} show that the number of ionising stars and the supernova rate depend very little on the metallicity at least between Z=0.008 and Z=0.04 (solar metallicity being Z=0.02). Hence, our assumption of solar metallicity for the clusters is well founded.    

The lower and upper mass limits in the IMF are respectively set to 1 and 100\,\msol, and supernovae occur for star masses $\ge$8\msol. Fig.~\ref{model} plots the Starburst99 predictions for the supernova rate (a) and emission rate of ionising photons (c). Fig.~\ref{model}b and Fig.~\ref{model}d show the non-thermal 6\,cm luminosity and the thermal 6\,cm luminosity, computed using formulae~(\ref{non thermal}) and~(\ref{thermal}), respectively. The considered mass for the cluster ($10^6$\,\msol) was chosen so as to predict radio luminosities which are of the order of those observed in the sources of NGC1365 and NGC1808 (see Table~\ref{table result}). Figs.~\ref{model}b and d show that after 3.5 Myr the non-thermal emission overcomes the thermal emission. This epoch depends mainly on the evolution of the thermal flux, which in turn depends on the slope of the IMF. For the chosen IMF, supernovae events stop at about $\sim 35$\,Myr after the birth of the cluster. The age of the radio sources in NGC1808 and NGC1365 should therefore lie between 3.5 and 35 Myr, given their fluxes and spectral indexes. How do the predicted cluster properties depend on the IMF cutoffs?

Decreasing the lower mass limit to 0.1\msol~would neither affect the ionising flux of the cluster nor its supernovae rate, but the mass of the cluster would be larger by a factor 2.5. On the contrary, an increase of the upper mass limit of the IMF will have little effect on the cluster masses, but will significantly increase the ionising luminosity of the cluster at very early stages (less than 3\,Myr). Yet, even though the supernovae occur earlier, at these very early stages the cm emission remains dominated by the thermal emission. Hence, a higher upper mass limit for the IMF does not significantly affect the dating of the clusters. For the clusters in NCG1808 and NGC1365, the cm spectral indexes show that the clusters are at a later evolutionary stage, more than 3\,Myr. Hence, a modification of the IMF upper mass would not greatly affect our conclusion. This is corroborated by a Starburst99 simulation, made with an upper mass cutoff of 150\,\msol. 

The subsequent step is to check if the observed MIR fluxes are consistent with those expected from a star cluster embedded in dust. Precise modelling would require detailed radiative transfer calculations. 
However, as our aim is to perform a check at the level of the order of magnitude only, we can simplify the problem and represent the cluster emission as a black-body at a given temperature. As mentioned earlier, the continuum from VSG observed by ISO in starburst galaxies can be well fit by black bodies at T$>$100\,K \citep{Laurent99}. We consider two values bracketing a plausible range of dust temperatures, 150\,K and 300\,K. Fig.~\ref{model}e plots the bolometric luminosity of the model cluster and Fig.~\ref{model}f translates it into the luminosity at 10.4\micro~assuming that the whole energy is reprocessed in a black-body at 150\,K (lower curve) and 300\,K (upper curve). The 10.4\micro~luminosity for T=300\,K is roughly one order of magnitude larger than for T=150\,K, and in both cases larger than the observed 10.4\micro~luminosity (Table~\ref{table result}). This difference is discussed in Sect.~\ref{colour-colour}. 

Overall, the interpretation that the MIR/radio sources observed in the starbursting environments of NGC1808 and NGC1365 are embedded young massive star clusters is satisfactory.  
\subsection{Dating and weighing the clusters}
\label{dating and weighing the clusters}
\begin{figure}[bp]
\begin{center}
\resizebox{9cm}{!}{\includegraphics*[scale=1.]{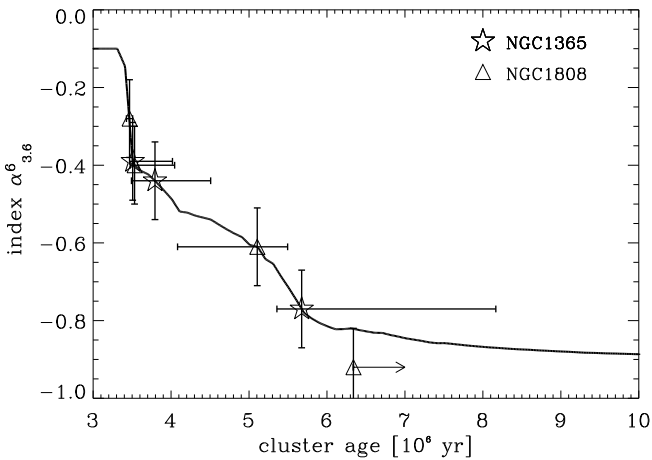}}
\caption{The curve shows the evolution with time of the spectral index $\alpha_{3.6}^{6}$ for our adopted cluster model (see Fig.~\ref{model} and corresponding description in Sect.~\ref{emission from young star clusters}). For a given model, an age may be attributed to every cluster in NGC1808 and NGC1365, depending only on the measured $\alpha^6_{3.6}$ value.}
\label{index}
\end{center}
\end{figure} 

Within the framework of the simple cluster modelling discussed above, it is possible to derive some quantitative parameters for each cluster. The index of the radio spectrum does not depend on the mass of the cluster whereas the supernova rate, Lyman photon emission rate and total luminosities are mass-scalable quantities. As illustrated in Fig.~\ref{index}, an age can be attributed to the embedded clusters by comparing the observed $\alpha$ value to the predicted one as the cluster evolves. Once an age is determined, we can compute the expected total 6cm flux for the $10^6$\,\msol~ cluster model. Rescaling this flux to the observed one provides the cluster mass. After mass and age have been derived for each cluster, it is possible to predict its 10.4\micro~flux if, as discussed above, we assume that the total luminosity from stars in the cluster is reemitted as a single black-body at temperature T. Results of these calculations are shown in Table~\ref{table result}. 

\begin{table*}[htbp]
\caption[]{Measurements and predictions of the properties of the embedded star clusters in NGC1808 and NGC1365.}
\begin{center}
\begin{tabular}{|l|cccc|ccc|} \hline
 & \multicolumn{4}{c|}{NGC1808} & \multicolumn{3}{c|}{NGC1365} \\
\multicolumn{1}{|c|}{Source name}  & M2 & M3 & M4 & M8 & M4 & M5 & M6 \\ \hline
$^a$Observed $\alpha^6_{3.6}$                      		  	     &       -0.92 &     -0.28 &     -0.40 &     -0.61 &     -0.39 &     -0.77 &     -0.44 	\\											       	
Estimated Age (Myr)             		   		  	     &  $\ge$ 6.3& 3.4$^{+0.1}_{-0.1}$ & 3.5$^{+0.5}_{-0.1}$ & 5.1$^{+0.4}_{-1.0}$ & 3.5$^{+0.5}_{-0.1}$ & 5.7$^{+2.5}_{-0.3}$ & 3.8$^{+0.7}_{-0.3}$ \\	       
$^a$Observed 6cm Flux [$\rm 10^{25} erg s^{-1} Hz^{-1}$]     		     &    9.2 & 15.2 & 14.6 & 14.6 &110.5 & 58.6 &113.4 \\																       
Predicted 6cm Flux [$\rm 10^{25} erg s^{-1} Hz^{-1}$]        		     &    9 &  57 &  64 &  41 &  64 &  37 &  53 \\																       
$^b$ Mass [$10^6\,$\msol]            		   		             &       1.6 &      0.3 &      0.3 &      0.4 &     1.7 &     1.6 &     2.1 \\												       
Observed 10.4\micro~Lum [$\rm 10^{26} erg s^{-1} Hz^{-1}$] 		     &   $\le$15 & 15 & $\le$15 & 20 & $\le$20 &250 &200 \\																       
$^c$Predicted 10.4\micro~Lum [$\rm 10^{26} erg s^{-1} Hz^{-1}$] (T=150\,K)   &    10 & 60 & 50 & 40 &360 &170 &390 \\															       
$^c$Predicted 10.4\micro~Lum [$\rm 10^{26} erg s^{-1} Hz^{-1}$] (T=300\,K)   &    50 & 370 & 300 & 270 &2300 &1100 &2500 \\ \hline                                                                                                   
 \end{tabular}
\end{center}
\label{table result}
$(^a)$ The radio measurements from \citet{Collison94} for NGC1808 and \citet{Sandqvist95} for NGC1365. The uncertainty on the spectral indexes is 0.1.

$(^b)$ Shifting the IMF lower mass cutoff to 0.1\msol, the mass would 2.5 times larger.

$(^c)$ The difference between observed and predicted luminosities at 10.4\micro~can be accounted for by absorption, the corresponding $A_V$ values are discussed in Sect.~\ref{colour-colour}.

\end{table*}

The final figures we obtain are similar for all the bright clusters (except NGC1808/M2). Their ages lie in the 3 to 6 Myr range. Even though their radio emission is not anymore dominated by the HII emission from massive stars, the clusters are very young. The masses inferred are between $3\,10^5$ and $2\,10^6$\msol, \textbf{comparable to typical globular cluster masses}.  

\subsection{Colour-colour diagrams}
\label{colour-colour}
\begin{figure*}[htbp]
\begin{center}
\resizebox{18cm}{!}{\includegraphics*[scale=1.]{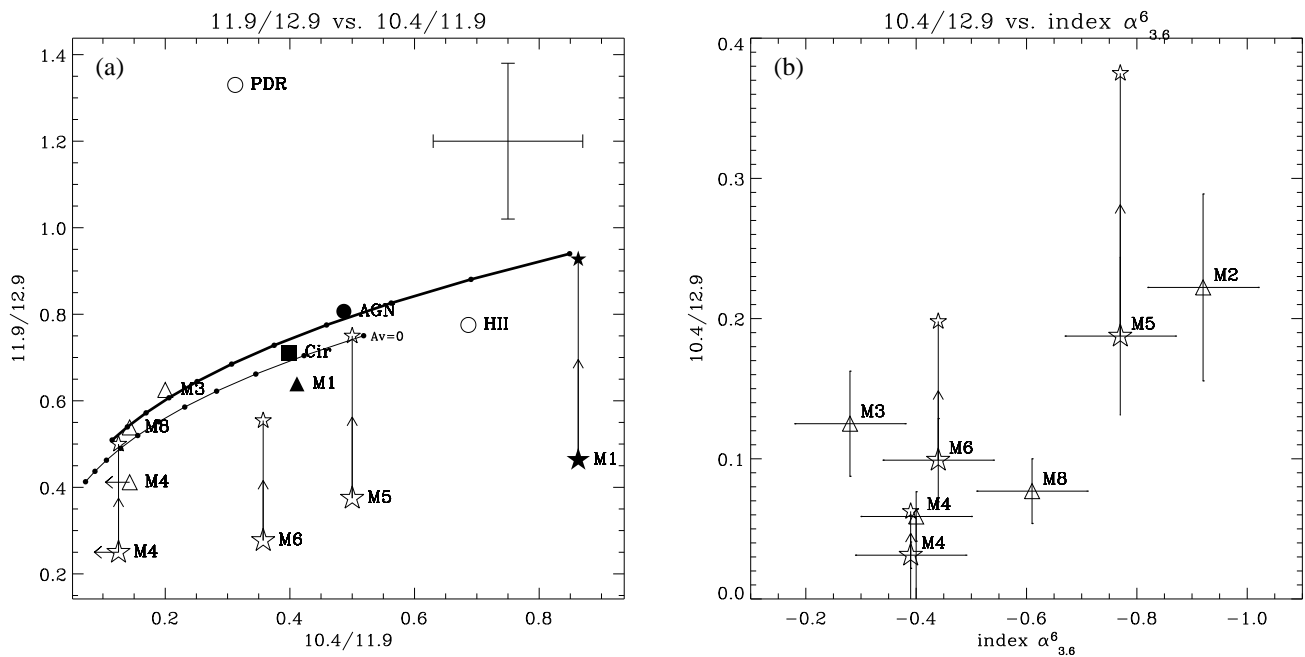}}
\caption{\textbf{(a):} The 11.9\micro~over 12.9\micro~flux ratio vs. the 10.4\micro~over 11.9\micro~flux ratio for the brightest sources in NGC1808 (triangles), NGC1365 (stars), Circinus (square) and the template spectra (circles). The filled symbol in each object represents the AGN. The solid lines show the location of black-bodies with temperature T=150\,K (thin line) and T=300\,K (thick line) with screen extinction ranging from $\rm A_V$=0 to 100 \citep[extinction law by][]{Dudley97}. The dots along the lines represent $\rm A_V$ steps of 10. The last dot to the right corresponds to $\rm A_V$=100. The error corresponding to our assumed 20$\%$ error on the photometric measurements leads to errors of 30$\%$ for the ratios: the typical error size is shown on the top left part of the plot. The small stars show the position that the sources in NGC1365 would occupy if we had overestimated the 12.9\micro~by a factor 2 (see Sect.~\ref{errors}). \textbf{(b):} The 10.4/12.9~flux ratio vs. the cm spectral index $\alpha$ for the clusters. For each point, the horizontal bar shows the $\pm0.1$ error on the spectral index, and the vertical bar shows the 30$\%$ error on the flux ratio. The small stars have the same meaning as in (a).}
\label{ratios}
\end{center}
\end{figure*} 
Modelling the MIR emission of the clusters with black-bodies was an easy and simple way to test this interpretation quantitatively. Yet, as shown with the MIR templates in Sect.~\ref{MIR emission}, emission in the N-band is expected to be more complex, with the presence of the silicate feature at 9.7\micro, PAH and emission lines ([SIV] 10.4\micro~and [NeII]12.8\micro). The 10.4\micro~filter includes the silicate feature and coincides with the [SIV] line, the 11.9\micro~filter coincides with a strong PAH band, and finally the 12.9\micro~filter samples the continuum outside the silicate feature and coincides with the 12.8\micro~[NeII] line. 

Fig.~\ref{ratios} presents two colour-colour diagrams: to the left the 11.9/12.9~flux ratio vs. the 10.4/11.9~flux ratio and to the right the 10.4/12.9~flux ratio vs. the cm spectral index $\alpha^6_{3.6}$.

Let us discuss first the left diagram, comparing the N-band colours of the brightest MIR/radio sources in NGC1808 and NGC1365 (and the AGN of Circinus) with the N-band colours of the template spectra of AGN, HII~and PDR (Sect.~\ref{MIR emission}). The filled symbols feature AGN while the open symbols are associated with the surrounding MIR/radio sources. On the colour-colour plot, the MIR/radio sources in NGC1808 cover a narrow range of 10.4/11.9 flux ratios. On the contrary, the MIR/radio sources in NGC1365 are spread in the 10.4/11.9 flux ratio and build up a sequence which is offset towards small 11.9/12.9 flux ratios. This offset could trace the uncertainty in the photometric calibration of the 12.9\micro~image due to local and unregistered variations of the sky transparency (see Sect.~\ref{errors}).  New observations are needed to confirm the 11.9/12.9 flux ratios in NGC1365. Even considering this uncertainty for NGC1365, the MIR colours of the clusters in both galaxies are redder than the templates. This could result from the steepness of the continuum slope between 10 and 14\micro, and/or from the presence of strong emission lines. A steep slope in this wavelength range can be produced by a deep 9.7\micro~silicate absorption feature. The colours of dust screened black-bodies at T=150\,K and 300\,K are overplotted in Fig.~\ref{ratios}a. Since the 10.4\micro~filter samples the deepest part of the 9.7\micro~silicate feature and the 12.9\micro~filter samples its outer edge, a large amount of extinction (several tens of magnitudes in the V-band) is consistent with the small 11.9/12.9 and 10.4/11.9 flux ratios observed (deep silicate absorption). We are aware that a dust screen model is a coarse approximation when the MIR continuum is emitted by the dust itself. Yet, it is useful to compare the measured 10.4\micro~fluxes to those predicted by the model discussed in Sect.~\ref{dating and weighing the clusters} and interpret the differences in terms of extinction by a dust screen. The $\rm A_V$ values derived this way range from a few magnitudes for T=150K to several tens of magnitudes ($\sim$50) for T=300K. PAH and/or [NeII]\,12.8\micro~lines are also expected to contribute in producing the large observed ratios.  Spectroscopic observations will be performed to investigate the silicate feature, the continuum slope and contribution of PAH and [NeII] lines. 

Fig.~\ref{ratios}b provides the 10.4/12.9~flux ratio vs. the cm spectral index $\alpha_{3.6}^{6}$. As seen in the previous paragraph, the 10.4/12.9 flux ratio decreases as the silicate absorption feature gets deeper, or as the [NeII] 12.8\micro~line gets more intense. The plot suggests a possible correlation between the 10.4/12.9~flux ratio and the spectral index. As shown in Sect.~\ref{dating and weighing the clusters}, the spectral index is expected to decrease with time. Thus the x-axis of Fig.~\ref{ratios}b may be regarded as a time indicator increasing from left to right. The increase of the 10.4/12.9~flux ratio together with that of the cm spectral index is broadly consistent with the picture of an evolving cluster, embedded in a cocoon of dust which dissipates with time. Along with the evolution of the cluster, 

(1) the ionising flux decreases as massive stars disappear through supernovae events, leading to a decrease of the [NeII] 12.8\micro~flux, 

(2) the dusty cocoon dissipates and consequently the dust column density decreases. This in turn translates into a weaker silicate absorption. 

Both effects result into an increase of the 10.4/12.9~flux ratio with time.

\section{Conclusion}
\label{conclusion}
New high angular resolution MIR observations have been presented for a set of nearby AGN, and one starburst galaxy. Of particular interest is a new population of MIR sources discovered in the central regions of the active galaxies NGC1808 and NGC1365 and found to be coincident with cm radio sources. These MIR/radio sources are interpreted in terms of young embedded star clusters. Considering the MIR narrow band colours and MIR flux densities, as well as the cm indexes and cm flux densities, a first-order model of these sources can be performed. From this analysis, we conclude that the MIR/radio sources are young (5\,Myr) globular-like ($10^6$\msol) clusters, still embedded in their dust cocoon. Further observations and modelling are being carried out to better understand their evolution. In particular, measurements of the ionised hydrogen line fluxes will help assessing the relative contributions of thermal and non-thermal radio emission. MIR spectroscopy will provide clues about the dust composition. Detailed modelling, including dust radiative transfer calculations, is being performed  (Galliano \& Delva, in preparation) and will be used to make new predictions about these exciting objects.

\label{conclusion}
\begin{acknowledgements}
We are gratefully indebted to the La Silla team operating the 3.6m telescope and the TIMMI2 instrument. We also acknowledge the pertinent comments from an anonymous referee as well as various discussions with ESO colleagues sharing the same scientific interests. 
\end{acknowledgements}

\bibliographystyle{aa}
\bibliography{manuref_link}

\end{document}